\documentclass[aps,pre,preprint,floatfix]{revtex4}%\documentclass[aps,pre,twocolumn,floatfix]{revtex4}
\usepackage{dcolumn}
\usepackage{graphicx}
\usepackage{amsmath}
\usepackage{amsfonts}
\usepackage{amssymb}
\usepackage{graphicx}
\usepackage{bm}
\usepackage{verbatim}

\newcommand{\bq}{\begin{equation}}
\newcommand{\eq}{\end{equation}}

\begin{document}
\title{
Modulated heat pulse propagation and partial transport barriers in chaotic magnetic fields}
\author{Diego del-Castillo-Negrete}
\author{Daniel Blazevski}
\affiliation{Oak Ridge National Laboratory \\ Oak Ridge, Tennessee 37831-8071, USA}
\begin{abstract}
Direct numerical simulations of the time dependent parallel heat transport equation modeling heat pulses driven by power modulation in 3-dimensional chaotic magnetic fields are presented. 
The numerical method is based on the Fourier formulation of a Lagrangian-Green's function method that provides an accurate and efficient technique for the solution of the parallel heat transport equation in the presence of harmonic power modulation.
The numerical results presented provide conclusive evidence that even in the absence of magnetic flux surfaces, chaotic magnetic field configurations with intermediate levels of stochasticity exhibit transport barriers to modulated heat pulse propagation.  In particular, high-order islands and remnants of destroyed flux surfaces (Cantori) act as partial  barriers that slow down or even stop the propagation of heat waves at places where the magnetic field connection length exhibits a strong gradient. 
Results on modulated heat pulse propagation in fully stochastic fields and across magnetic islands are also presented. 
In qualitative agreement with recent experiments in LHD and DIII-D,
it is shown that the elliptic (O) and hyperbolic (X) points of magnetic islands have a direct impact on the spatio-temporal dependence of the amplitude of modulated heat pulses. 
\end{abstract}
\maketitle

%%%%%%%%%%%%%%%%%%%%%%%%%%%%%%
\section{Introduction}
%%%%%%%%%%%%%%%%%%%%%%%%%%%%%%

Magnetic field topology in general, and magnetic field line chaos in particular play a key role in the confinement of fusion plasmas. Ideally, magnetic configuration should consists of nested flux surfaces foliated by non-chaotic (integrable) field lines. However, in practice, imperfections  in the coils as well as spontaneously generated and externally induced magneto-hydrodynamic instabilities typically lead to configuration consisting of a complex fractal mixture of integrable and chaotic field lines. Although it is widely recognized that magnetic field stochasticity is closely linked to lack of confinement, the precise quantitative connection is highly non-trivial and not fully-understood.  This is because magnetic field chaos by itself is a highly nontrivial nonlinear dynamics problem, but more importantly, because the connection between  the dynamics and topology of the magnetic field and the transport of say density and temperature is not as straightforward as one might naively expect.  A problem of particular interest is how different levels of magnetic stochasticty manifest on transport. Between the two extreme limits of nested flux surfaces and full magnetic field stochasticity, there is a vast regime of weak and intermediate chaos in which the magnetic field geometry exhibits fractal behavior and the conventional transport models based on simplified assumptions (e.g., quasilinear diffusion and Gaussian random-walk type models) break-down. The present paper focusses for the most part on this regime in the context of electron heat transport. 

The highly nontrivial role of the magnetic field stochasticity on electron heat transport stems from the strong asymmetry between the parallel (with respect to the local magnetic vector field) and the perpendicular heat conductivities. That is, electron heat transport is inherently anisotropic and the geometry of this anisotropy is dictated by the dynamics of the magnetic field. When the magnetic field is fully integrable one can cope with this asymmetry by adopting an appropriate coordinate system. However, in the case of general three-dimensional chaotic fields this is not possible. As a result, electron heat transport in chaotic magnetic fields is both a numerically challenging problem and a fascinating  physics problem with highly nontrivial consequences (most of them yet to be explored) to the magnetic confinement of fusion plasmas. 

In previous publications we considered some aspects of this problem including a study of the break-down of the standard diffusive model in the case of full magnetic field stochasticity \cite{DL,DL_pop} and the role of partial transport barriers in monotonic and reversed shear configurations \cite{dan_diego_2013,diego_dan_2014}. Going beyond these works, in the present paper we study electron heat pulse propagation driven by power modulation in three dimensional chaotic magnetic fields. 

Understanding heat pulse  transport due to power modulation is important because transport coefficients are often estimated from perturbative experiments. In particular,  effective diffusivities, advection velocities, and damping rates can be inferred from local measurements of  the amplitude and the phase of the propagation of harmonic temperature perturbations in plasmas as discussed in  
Ref.~\cite{jacchia_eta_al_Phys_Fluids_1991} and more recently in 
Ref.~\cite{berkel_et_al_PoP_2014}. For a review of perturbative experiments see 
Ref.~\cite{Cardozo_1995}. Heat pulse propagation has also been used to study the  connection  between transport and bifurcations of magnetic field  in modulated electron cyclotron heating perturbative experiments in the Large Helical Device   (LHD) \cite{ida_LHD_2013} and more recently in the DIII-D tokamak \cite{Ida_etal_2015}.  In these works it  was shown that amplitude and time-delayed measurements of modulated heat pulses can be used to infer changes in the magnetic field topology. 
Most of the published material in this area has been experimental and to the best of our knowledge the present paper is the first theoretical study of heat pulse  transport due to power modulation in the presence of chaotic magnetic fields based on the numerical solution of the parallel electron heat transport equation. 

Our main goal is to present a detailed study of the propagation of electron heat waves driven by power modulation in plasmas with different levels of magnetic field stochasticity. In all the cases studied there are no magnetic flux surfaces. That is, from the point of view of the magnetic field there is no confinement. However, one of the main conclusions of the present paper is that despite this, partial barriers to heat wave propagation form in regions where the magnetic field connection length exhibits strong gradients. 
We refer to these barriers as ``partial" in the sense that heat can leak through them but, depending on the magnetic field connection length and the length scale $1/\gamma=\sqrt{2 \chi_\parallel/\omega}$ (where  $\chi_\parallel$ is the parallel diffusivity and $\omega$ is the power modulation frequency), the flux can be very small. For large  frequencies or small parallel thermal conductivities, parallel heat transport is strongly damped and the magnetic field  partial barriers act as robust  barriers to radial heat transport. As a result, in this case the radial amplitude of the heat wave vanishes and its phase speed slows down to a halt at the location of the partial barrier. 
On the other hand, in the limit of small $\gamma$, that is, for small modulation frequencies or large parallel conductivities, parallel heat transport is largely unimpeded and global radial heat transport observed. In this case, the radial amplitude and phase speed of the heat wave remain finite.

We also consider heat wave propagation in the case of very strong magnetic field stochasticity. This regime, which is characterized by the complete lack of magnetic islands, manifest itself in Poincare plots as a seemingly featureless homogeneous distribution of points. It is shown that in this extreme stochastic regime heat wave propagation follows diffusive transport for high enough  modulation frequencies (i.e., large values of $\gamma$). 
To conclude, motivated by the experimental study reported in \cite{Ida_etal_2015}, we present preliminary results on modulated heat pulse propagation  across magnetic islands. The geometry used in this calculation is toroidal and the magnetic field was obtained using the magneto-hydrodynamic equilibrium code SIESTA \cite{hirshman_etal_2011}. The main objective of these calculations is to explore the impact of magnetic islands on the propagation of heat waves driven by power modulation. 

As mentioned before, beyond its theoretical and practical relevance, the study of electron heat transport in stochastic magnetic field is interesting from the computational physics perspective due to the uniquely challenging numerical setting of the problem. In particular, as it has been discussed in previous publications, the extreme anisotropy and the intrinsic 3-dimensional nature of the problem in the case of chaotic fields posses serious problem to standard (e.g., grid based continuum methods) numerical algorithms. In fusion plasmas, the ratio of the parallel and perpendicular electron thermal conductivities, $\chi_{||}/\chi_\perp$, might exceed $10^{10}$ \cite{Braginskii,Fitzpatrick}. Motivated by this, here we focus on the study of purely parallel transport, i.e. we assume $\chi_\perp=0$. To address this transport regime, we use the Lagrangian-Greene's function (LG) method \cite{DL,DL_pop} that provides an efficient and accurate technique to solve the parallel heat transport equation.  Although the original formulation of the LG method contemplated sources which in principle can have arbitrary spatio-temporal dependences, the present work is the first to present specific numerical results for time periodic sources.  To deal with this problem, here we present a novel reformulation of the LG method in Fourier space that  is significantly more efficient than the original implementation in real space.  

The rest of the paper is organized as follows. Section~\ref{LG_review} present the LG method in Fourier space, and Sec.~\ref{models} discusses the magnetic field models used in the transport calculations. 
The main numerical results are presented in Sec.~\ref{results} which consists of three subsection. Subsection~\ref{ptb_results} deals with the problem of modulated heat pulse propagation in the presence of partial transport barriers in magnetic fields with weak and moderate levels of stochasticity. Subsection~\ref{chaos_results}  considers the case of strong magnetic field stochasticity, and 
Subsection~\ref{island_results} studies heat pulse propagation in the presence of magnetic islands in toroidal geometry. 
The conclusions along with a brief summary of the results are presented in Section~\ref{conclusions}.

%%%%%%%%%%%%%%%%%%%%%%%%%%%%%%
\section{Lagrangian-Green's function method for modulated heat transport}
\label{LG_review}
%%%%%%%%%%%%%%%%%%%%%%%%%%%%%%

Our starting point for the modeling of the propagation of modulated heat pulses is the electron heat transport equation 
$(3/2) \partial_t (n T) = - \nabla \cdot \mathbf{q} + S$,
where $T$ is the plasma temperature,  $n$ is the plasma density, $S$ denotes a source, and $\mathbf{q}$ is the heat flux consisting of a parallel (along the magnetic field) and a perpendicular 
component, ${\bf q}= q_\parallel {\bf \hat b} + {\bf q}_\perp$, where ${\bf \hat b}={\bf B}/|B|$ is the unit magnetic field vector. 
The source is assumed to consists of a steady part, $S_0({\bf r})$, maintaining the equilibrium temperature, $T_0({\bf r})$, and a 
time-dependent part, $S_1({\bf r},t)$, accounting for the power modulation. As it is customary done in perturbative transport studies, we write $T=T_0({\bf r})+T_1({\bf r},t)$ where $T_1({\bf r},t)$ denotes the response to the modulation and consider the transport equation 
\begin{equation}
\label{eq_II_1}
\frac{3}{2} n_0 \frac{\partial  T_1}{\partial t} = - \nabla \cdot \mathbf{q}_1 + S_1({\bf r},t) \, ,
\end{equation}
where for simplicity we assume a constant plasma density, $n_0$. 

As mentioned in the introduction, we will limit  attention to parallel heat transport in the extreme
anisotropic regime 
i.e., we will assume ${{\bf q}_1}_{\perp}=0$. 
As it is well-known, this  regime is numerically challenging and it calls for the use of advanced numerical algorithms able to
accurately and efficiently cope with extreme anisotropy. This is particularly demanding when the magnetic field exhibit stochasticity which is the case of the present study. 
To address this problem, here we use a Lagrangian-Green's function (LG) method. The LG method was originally proposed in Ref.~\cite{DL} for the case of parallel heat transport ($\chi_\perp=0$), and extended to the  $\chi_\perp \neq 0$ case in Ref.~\cite{luis_jcp_2013}.  In previous publications, this method was applied to study transport in different magnetic field configurations \cite{DL_pop,dan_diego_2013,diego_dan_2014} but limiting attention to cases in which the heat source was either absent or time independent. Going beyond these previous studies,  we consider heat transport in the presence of time modulated  sources, and present in this section a re-formulation of the LG method in Fourier space to deal with this  problem efficiently.  
 
Since the Lagrangian-Green's function (LG) method in real space has been discussed in  detail in
previous publications (see Refs.~\cite{DL,DL_pop,luis_jcp_2013}) here we limit to a brief recapitulation of the main ideas. 
Given an initial temperature distribution $T_{t=0}({\bf r})=T({\bf r},t=0)$, 
and a source $S({\bf r},t)$, the temperature at a given point in space ${\bf r}_0$, at a time $t$, is obtained by summing all the contributions of the initial condition and the source along the unique  magnetic field path going through ${\bf r}_0$. That is, 
\bq
\label{eq_II_10}
T({\bf r}_0,t) = \int_{-\infty}^\infty T_{t=0} \left[ {\bf r}(s') \right] {\cal G}(s',t) ds' +
 \int_0^t dt'\, \int_{-\infty}^\infty ds' S \left[ {\bf r}(s'),t' \right] {\cal G}(s',t-t') \, ,
\eq
where 
${\bf r}={\bf r}(s)$ denotes the magnetic field line parametrized using the arc length $s$ and obtained from the solution of the 
initial value problem 
\begin{equation}
\label{eq_II_11}
\frac{d \mathbf{r}}{d s} = \mathbf{\hat{b}}\, , \qquad \mathbf{r}(0) = \mathbf{r}_0 \, .
\end{equation}
Note that, to ease the notation, we have rescaled the variables to absorb the constant $3n_0/2$ and have dropped the subindex ``$1$" with the understanding that from now on, $T$ denotes the  response to the 
time dependent power modulation.  

The implementation of the LG method
requires three elements: an ODE integrator for solving the field line trajectories in 
Eq.~(\ref{eq_II_11}), an
interpolation method for the function $T_{t=0}({\bf r})$ on the field line, and a 
numerical quadrature to evaluate the Green's function integrals in Eq.~(\ref{eq_II_10}). 
These elements are relatively straightforward to implement numerically,  making the LG  algorithm a versatile, efficient, and accurate  method for the computation of heat transport in magnetized plasmas. By construction, the LG method preserves the positivity of the temperature evolution, and avoids completely the
pollution issues encountered in finite difference and finite elements algorithms. Also, because of the parallel
nature of the Lagrangian calculation, the formulation naturally leads
to a massively parallel implementation. In particular, 
the computation of  $T$ at  ${\bf r}_0$ at time $t$
does not require the computation of $T$  in the neighborhood of ${\bf r}_0$ or the computation of $T$ at previous times, as  it is the case in finite different methods.  Further details on the method and the numerical implementation can be found in Refs.~\cite{DL_pop,luis_jcp_2013}.

In principle, Eq.~(\ref{eq_II_10}) provides the solution of the parallel heat transport problem for an arbitrary initial condition and source. However, in the case of time dependent sources this formulation might be  time consuming due to the need to evaluate for each $t$ and ${\bf r}_0$ a double integral, one in time and another along the field line, in the second term on the right hand side of Eq.~(\ref{eq_II_10}).  However, in the case of time periodic sources the computation can be significantly accelerated by working in Fourier space. 

Taking the Fourier transform in time, $\tilde{T}(s,\omega)=\int_{-\infty}^\infty T(s,t) e^{-i \omega t} dt$,
of the parallel heat  transport along a magnetic field line,
\begin{equation}
\label{eq_II_6}
\partial_t T =  \chi_\parallel \partial^2_s T + S(s,t) \, ,
\end{equation}
we have
\bq
\label{eq_fourier_1}
\frac{d^2 \tilde{T}}{dz^2}-i  \tilde{T} =  \tilde{P} \, ,
\eq
where $z=\sqrt{\omega/\chi_\parallel}\, s$ is a rescaled coordinated along the magnetic field line, 
$\tilde{P}=-\tilde{S}(s,\omega)/\omega$, and it is assumed that $\omega >0$. Note that the solution for $\omega <0$ can be obtained from 
$\tilde{T}(s,-\omega)=\tilde{T}^*(s,\omega)$ which follows from the fact that $T(s,t)$ is real. 
The solution of Eq.~(\ref{eq_fourier_1}) in an unbounded domain is 
\bq
\tilde{T}(z,\omega)=\int_{-\infty}^\infty dz' \tilde{P}(z',\omega) G\left(z-z' \right) \, ,
\eq
where the corresponding Green's function, $G\left(z-z' \right)$, is 
\bq
G=\frac{-1}{\sqrt{2} (1+i)} \exp\left[  -\frac{1}{\sqrt{2}} (1+i) \left| z- z'\right| \right] \, .
\eq
From here it follows that
\bq
\label{green_sol_fourier}
\tilde{T}(s,\omega)=\frac{1}{\sqrt{2 \omega \chi_\parallel}} \frac{1}{(1+i)}
\int_{-\infty}^\infty ds' \tilde{S}(s',\omega)  \exp\left[  -(1+i) \sqrt{\frac{\omega}{2\chi_\parallel}} \left| s- s'\right| \right]
\, .
\eq
This Green's function solution along a magnetic field line is the main building block to construct the 
solution of the heat transport equation at an arbitrary point in space. In particular, the solution of the heat transport 
equation in Eq.~(\ref{eq_II_1}) with $T({\bf r},t=0)=0$ at  ${\bf r}_0$  is  given by 
\bq
\label{green_sol_fourier}
\tilde{T}({\bf r}_0,\omega)=\frac{1}{\sqrt{2 \omega \chi_\parallel}} \frac{1}{(1+i)}
\int_{-\infty}^\infty ds' \tilde{S}[{\bf r}(s'),\omega]  \exp\left[  -(1+i) \sqrt{\frac{\omega}{2\chi_\parallel}} \left|s'\right| \right]
\, ,
\eq
where ${\bf r}(s)$ is the solution of the magnetic field line Eq.~(\ref{eq_II_11}) with initial condition ${\bf r}(s=0)={\bf r}_0$. 

In the case of a source with a separable spatio-temporal dependence, 
$S({\bf r},t)=Q({\bf r}) \Lambda (t)$,
we have
\bq
\label{green_sol_fourier_sep}
\tilde{T}({\bf r}_0,\omega)=\frac{1}{\sqrt{2 \omega \chi_\parallel}} \frac{\tilde{\Lambda}(\omega)}{(1+i)}
\int_{-\infty}^\infty ds'  Q[{\bf r}(s')]  \exp\left[  -(1+i) \sqrt{\frac{\omega}{2\chi_\parallel}} \left|s'\right| \right]
\, .
\eq
When the time dependence is periodic with fundamental frequency $\omega_0$,
\bq
\Lambda(t)=\sum_{m=-\infty}^\infty \Lambda_m e^{i m \omega_0 t} \, ,
\eq
with $\Lambda_m^*=\Lambda_{-m}$ and $\Lambda_{0}=0$,
substituting $\tilde{\Lambda}(\omega)=2 \pi \sum_m \Lambda_m \delta (m\omega_0-\omega)$ into Eq.~(\ref{green_sol_fourier_sep}) leads to the Lagrangian-Green's function solution at ${\bf r}_0$ of  the anisotropic heat transport equation driven by a periodic time dependent source, 
\bq
\label{LG-F}
T({\bf r}_0,t)= {\rm Re} \sum_{m=1}^\infty \frac{\Lambda_m \, {\cal I}_m}{\sqrt{m} \chi_\parallel \left(1 + i\right) \gamma } e^{i m \omega_0 t} \, , 
\qquad
{\cal I}_m = \int_{-\infty}^\infty ds'\, Q\left[ {\bf r}(s') \right] e^{-\gamma \sqrt{m} \left(1 + i \right) |s'|} \, ,
\eq
where ${\rm Re}$ denotes the real part and
\bq
\label{gamma_def}
\gamma=\sqrt{\frac{\omega_0}{2 \chi_\parallel}} \, .
\eq
Note that the evaluation of Eq.~(\ref{LG-F}) requires only one integral along the field line for each spatial evaluation point 
${\bf r}_0$, a significant simplification compared to the solution in Eq.~(\ref{eq_II_10}) requiring a double integral in space and time.  

Like in the standard LG method, the numerical computation requires the truncation of the integral along the field line in Eq.~(\ref{LG-F}). 
Introducing a cut-off $s_f$, we get the following expressions for the numerical computation of the real and imaginary parts of 
of ${{\cal I}_m}$
\bq
\label{I_app_r}
{{\cal I}_m}_R = \int_{-s_f}^{s_f} ds'\, Q\left[ {\bf r}(s') \right] e^{-\gamma \sqrt{m} |s'|} \cos \left(\gamma \sqrt{m} |s'| \right)
\eq
\bq
\label{I_app_i}
{{\cal I}_m}_I =- \int_{-s_f}^{s_f}  ds'\, Q\left[ {\bf r}(s') \right] e^{-\gamma \sqrt{m} |s'|} \sin \left(\gamma \sqrt{m} |s'| \right) \, .
\eq
To compute these integrals, we formulate them as two ordinary differential equations and solve them simultaneously with the forward and backward  magnetic field line orbits. That is, we write
\bq
{{\cal I}_m}_R = A(s_f) \, , \qquad {{\cal I}_m}_I =  B(s_f)
\eq
where $A(s_f)$and $B(s_f)$ are the solutions of the ordinary differential equations 
\bq
\label{ode_1}
\frac{d A}{ds}=\left[ Q\left({\bf r}_-(s)\right) + Q\left({\bf r}_+(s)\right) \right]  e^{-\gamma \sqrt{m} s} \cos \left(\gamma \sqrt{m} s\right) \, , 
\eq
\bq
\label{ode_2}
\frac{d B}{ds}=\left[ Q\left({\bf r}_-(s)\right) + Q\left({\bf r}_+(s)\right) \right]  e^{-\gamma \sqrt{m} s} \sin \left(\gamma \sqrt{m} s \right) \, , 
\eq
at $s=s_f$ with initial conditions 
\bq
A(s=0)=0 \, , \qquad B(s=0)=0 \, ,
\eq
and ${\bf r}_-(s)$ and ${\bf r}_+(s)$ denote the backward and forward magnetic field line orbits defined by the ordinary differential equations
\bq
\frac{d {\bf r}_-}{ds}= -\hat{\bf{b}}(s) \, , 
\qquad 
\frac{d {\bf r}_+}{ds}= \hat{\bf{b}}(s) \, , 
\eq
with initial conditions
\bq
{\bf r_-}(0)={\bf r_0} \, , \qquad {\bf r_+}(0)={\bf r_0} \, .
\eq
The cut-off parameter $s_f$, in the integrals in Eqs.~(\ref{I_app_r})-(\ref{I_app_i}), which defines the integration interval for the ordinary differential equations in (\ref{ode_1})-(\ref{ode_2}), is chosen large enough so that the truncation errors are kept below a prescribed tolerance. 

%%%%%%%%%%%%%%%%%%%%%%%%%%%%%%%%%%%%%%%%
\section{Magnetic field model}
\label{models}
%%%%%%%%%%%%%%%%%%%%%%%%%%%%%%%%%%%%%%%%

In the numerical solution of the parallel heat transport equation we consider two classes of magnetic fields. 
For the study of the role of magnetic field stochasticity on the propagation of modulated heat pulses we assume a cylindrical geometry and consider a magnetic field consisting of an axisymmetric helical field with a monotonic $q$ profile perturbed by a linear superposition of modes with amplitude $\epsilon$. The main advantage of this analytical model is that the amount of stochasticity can be  controlled with the single parameter 
$\epsilon$. Also, in this model   the location  of the magnetic islands and the partial transport barriers can be controlled with the $q$ profile and the poloidal and toroidal mode numbers of the perturbation. 
For the study of the propagation of heat pulses across magnetic islands in toroidal geometry we use the equilibrium magneto-hydrodynamics code SIESTA. In this section we focus on the description of the analytical model in cylindrical geometry. 

The magnetic field is given by 
\begin{equation}
\label{BB_model}
 \mathbf{B}(r, \theta, z) = \mathbf{B}_0 (r) + \mathbf{B}_1(r, \theta, z) \, ,
\end{equation}
where $(r,\theta,z)$ denote cylindrical coordinates, ${\bf B}_0(r)$ is a helical
field of the form, 
\bq
\label{B_model}
{\bf B}_0=B_\theta(r) \,  \hat{\bf e}_\theta + B_z \hat{\bf e}_z \, , 
\eq
with $B_z$ constant,
and $\mathbf{B}_1(r, \theta, z)$  is a perturbation. 
The period of the 
cylindrical domain is $L=2 \pi R$, i.e. $\mathbf{B}(r, \theta, z+L)=\mathbf{B}(r, \theta, z)$. In  all the calculations we assume $R=5$ and $B_z=1$.
The shear of the helical magnetic  field, i.e. the dependence of the azimuthal rotation of the field as function of the radius, is determined by the safety factor $q(r)=r B_z/(R B_\theta)$. Here we assume 
\bq
\label{B_theta}
B_\theta(r)=\frac{B (r /\lambda)}{1+(r/\lambda)^2} \, ,
\eq
with $B=0.1$ and $\lambda=0.573$. This implies
\bq
q(r)=q_0\left( 1 - \frac{r^2}{\lambda^2}\right )\, ,
\eq
which is an increasing linear function in the radial flux variable
\bq
\label{psi}
R^2 \psi= \frac{r^2}{2} \, . 
\eq

It is assumed that the  perturbation, ${\bf B}_1$, has no $z$-component 
and given by
 \bq
 \label{perturbation_B}
{\bf B}_1=\nabla \times \left[ A_z(r,\theta,z) \hat{\bf e}_z \right] \, ,
\eq
where the magnetic potential is given by a superposition of normal modes of the form
\bq
\label{perturbation}
A_z=\sum_{m,n} A_{mn}(r)  \cos \left( m \theta - n z/R  + \zeta_{mn}  \right) \, ,
\eq
where $\zeta_{mn}$ are constant phases.
The mode amplitude functions are given by 
\bq
\label{eq_21}
A_{mn}(r)= \epsilon a(r) \left( \frac{r}{r_*} \right)^m \exp \left[
  \left( \frac{r_*-r_0}{\sqrt{2} \sigma}\right)^2 -\left(
  \frac{r-r_0}{\sqrt{2} \sigma}\right)^2 \right] \, , 
\eq with 
\bq
r_*=\lambda \sqrt{\frac{m}{n} \left( \frac{B R}{B_z \lambda}\right)-1}
\, , \qquad r_0=r_*-\frac{m \sigma^2}{r_*} \, .  
\eq 
By construction, $r=r_*$ corresponds to the location of the $(m,n)$
resonance, i.e., $q(r=r_*)=r_*
B_z/(R B_\theta(r_*))=m/n$.  The value of $r_0$ is chosen to guarantee
that at the resonance, the perturbation reaches a maximum, 
$dA_{mn}/dr (r=r_*)=0$, given by $A_{mn}(r_*)=\epsilon$. The prefactor
$(r/r_*)^m$ is included to guarantee the regularity of the radial
eigenfunction near the origin, $r\sim 0$ for the given $m$. Finally,
the function, 
\bq
\label{a_fcn}
a(r)=\frac{1}{2} \left [
1 - \tanh \left( 
\frac{r-1}{\ell}
\right )
\right ] \, ,
\eq
is introduced to guarantee
the vanishing of the perturbation for $r=1$.  
In all calculations $\epsilon\sim {\cal O}\left(10^{-4}\right)$,
which, consistent with the tokamak ordering approximation, implies that
the numerically computed variation of the field is of the order of
 $\left | \partial_s B / B \right | \sim 10^{-4}$. 
This magnetic field model is the same as the one used in the calculation reported in Ref.~\cite{diego_dan_2014} where a 
study of the the decay of heat pulses in the absence of power modulation was presented.  

%%%%%%%%%%%%%%%%%%%%%%%%%%%%%%%%%%%%%%%%
\section{Numerical results}
\label{results}
%%%%%%%%%%%%%%%%%%%%%%%%%%%%%%%%%%%%%%%%

This section presents the main numerical results on the study of modulated heat pulse propagation for different levels of magnetic field stochasticity. 

%%%%%%%%%%%%%%%%%%%%%%%%%%%%%%%%%%%%%%%%
\subsection{Partial transport barriers in weak and moderate chaotic magnetic fields}
\label{ptb_results}
%%%%%%%%%%%%%%%%%%%%%%%%%%%%%%%%%%%%%%%%

To study the role of weak and moderate levels of magnetic field stochasticity on transport we considered 
a perturbation consisting of four overlapping modes in Eq.~(\ref{perturbation}) with
$(m,n)=\left \{  (11, 6), (5, 2), (9, 2), (4, 1) \right\}$
and the following values of the perturbation amplitude: $\epsilon=1.5 \times 10^{-4}$, $\epsilon=2.0 \times 10^{-4}$ , $\epsilon=2.5 \times 10^{-4}$  and $\epsilon=3.0 \times 10^{-4}$. 
For comparison purposes, it  is worth noting that this is the same magnetic field perturbation used in 
 in the study of {\em decaying} heat pulses, i.e. without power modulation, in Ref.~\cite{diego_dan_2014}.

We assume a separable source of the form, $S=R({\bf r}) \Lambda(t)$, 
with $R(\bf{r})$ a  radially localized function in space and $\Lambda(t)$ a single harmonic periodic function of time of the form
\bq
\label{source_rt}
R({\bf r}) = \exp \left [ - \frac{\left( R^2 \psi - R^2 \psi_0 \right)^2}{\sigma^2}\right ] \, , \qquad
\Lambda(t)= e^{i \omega_0 t} \, ,
\eq
where $\psi$ denotes the radial flux variable in Eq.~(\ref{psi}). In all the simulations presented in this paper, 
$R^2 \psi_0=0.325$ and $\sigma=0.008$. 

\begin{figure}
\includegraphics[width=0.55\columnwidth]{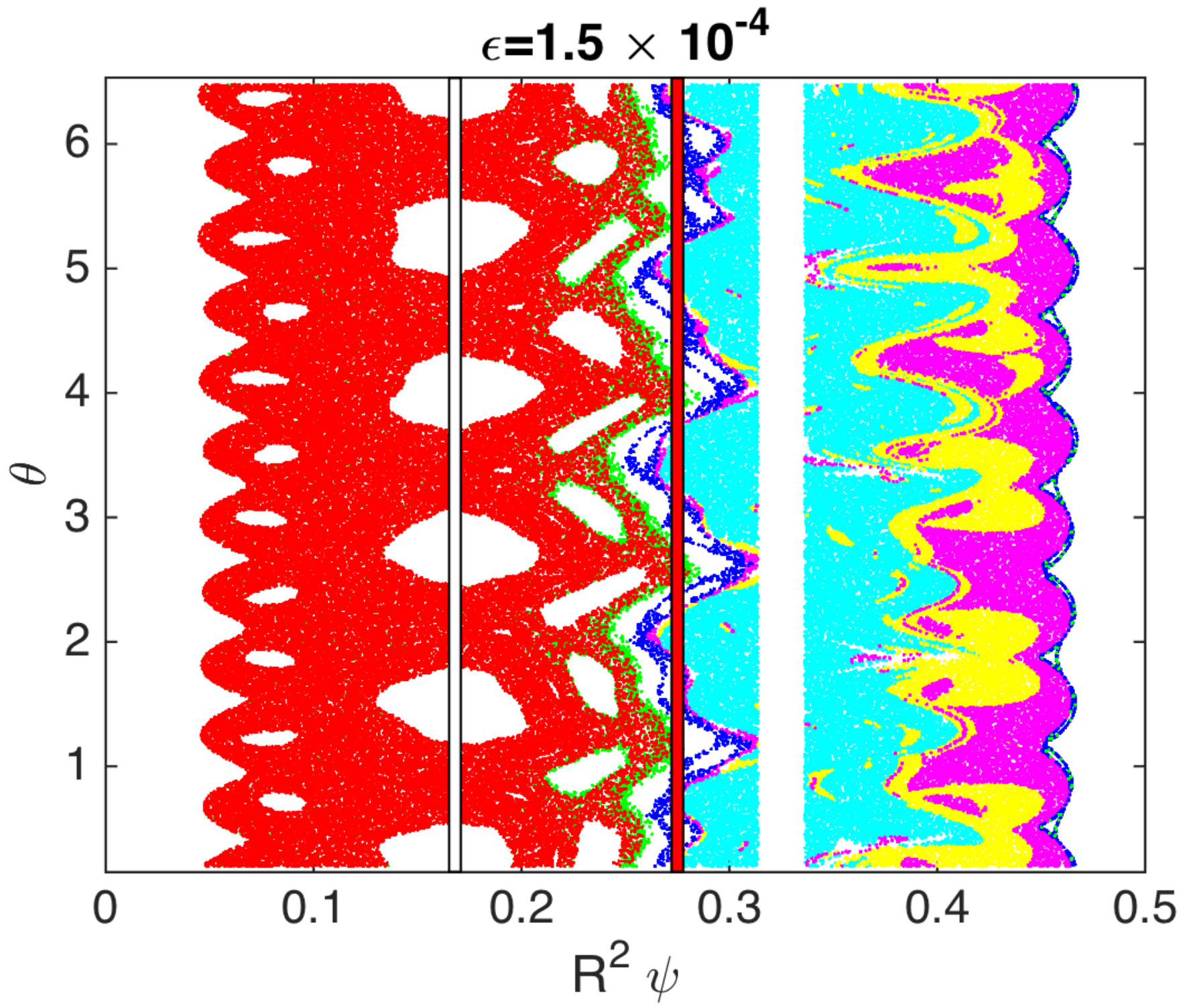}
\includegraphics[width=0.55\columnwidth]{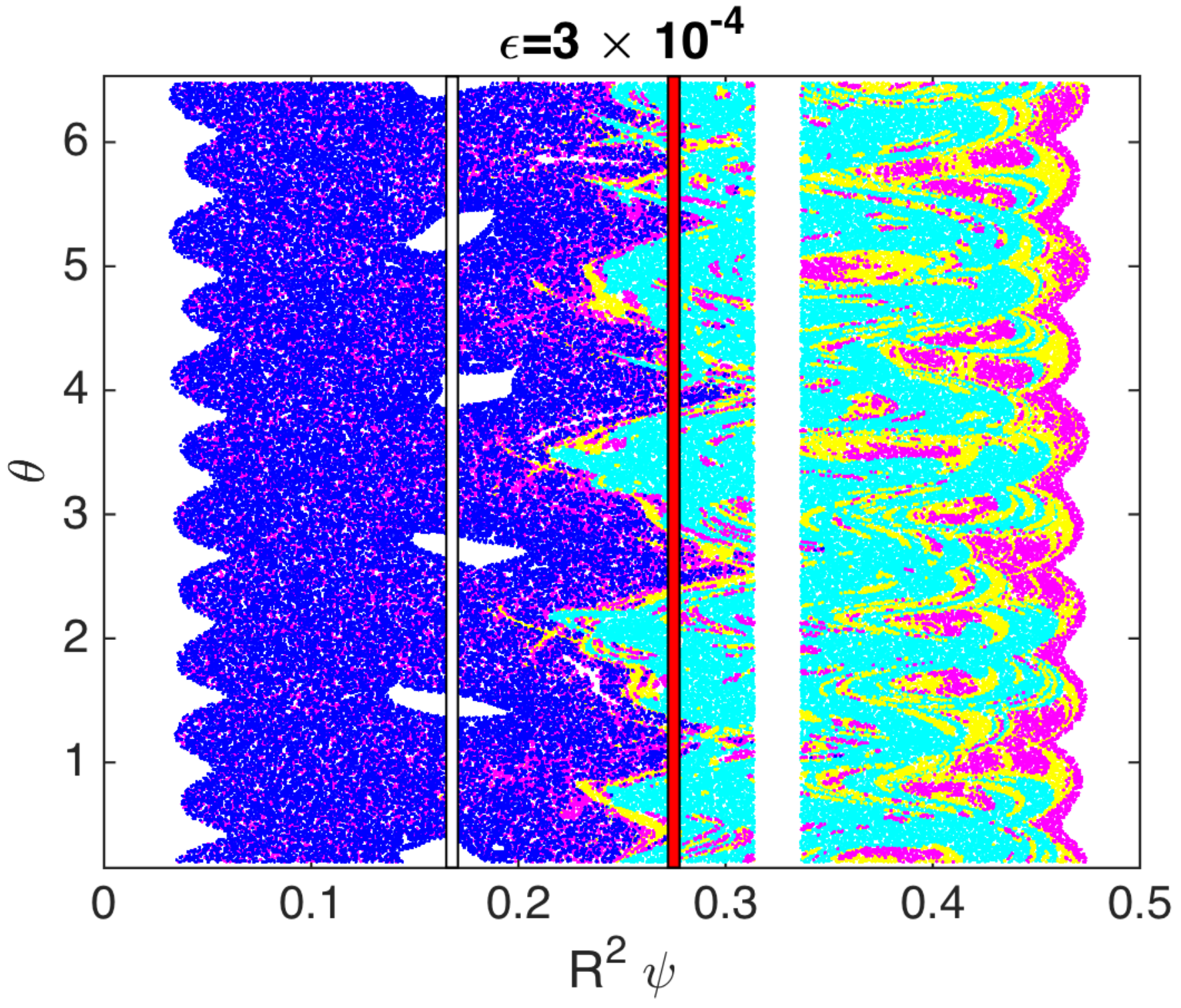}
\caption{(Color online) 
\label{poinc_connection_fig}
Magnetic field connection length. The figures show Poincare plots of the
magnetic field  in Eqs.~(\ref{BB_model}) perturbed by the four modes (see  
Sec.~\ref{ptb_results}) with amplitudes $\epsilon=1.5 \times 10^{-4}$ (low stochasticity) and 
$\epsilon=3.0 \times 10^{-4}$ (moderate stochasticity).
The strip centered at $R^2 \psi =0.325$ indicates the location of the source in Eq.~(\ref{source_rt}), the 
line at $R^2 \psi =0.275$ denotes the location of the partial barrier, and 
the line at $R^2 \psi = 0.168$ denotes the location of the main $(5,2)$ resonance.
In each plot the points correspond to a {\em single} initial condition indicating absence of magnetic flux surfaces in 
$R^2 \psi \in (0.05, 0.475)$. The points are color coded by the value of magnetic field connection length, $\ell_B$, defined in Eq.(\ref{cl}). 
$\ell_B<5 $ (Cyan); $5<\ell_B<10 $ (Yellow); $10<\ell_B<10^2 $ (Magenta);
$10^2 <\ell_B< 10^3 $ (Blue); $10^3 < \ell_B<10^4 $ (Green);
$10^4 < \ell_B$ (Red). [Adapted from Ref.~\cite{diego_dan_2014}]}
\end{figure}

For all the values of $\epsilon$ considered, the magnetic field configuration  has no flux surfaces in the computational domain
$\psi \in (0.075, 0.45)$.
This is clearly observed in Fig.~\ref{poinc_connection_fig} that shows Poincare sections obtained from 
the numerical integration of a {\em single}  magnetic field initial condition for a large number of crossings with
$\epsilon=1.5 \times 10^{-4}$ and 
$\epsilon=3.0 \times 10^{-4}$. 
Similar results (see Ref.~\cite{diego_dan_2014}) are obtained for the intermediate values $\epsilon=2.0 \times 10^{-4}$ and  $\epsilon=2.5 \times 10^{-4}$. 

%%%%%%%%%%%%%%%%%%%%%%%%%%%%%%%%%%%%
%%%%%%%%%%%%%%%%%%%%%%%%%%%%%%%%%%%%

The Poincare plots in Fig.~\ref{poinc_connection_fig} are color coded by the magnetic field connection length, $\ell_{B}$, which following Ref.~\cite{diego_dan_2014} we define as follows.  
Given a region, ${\cal R}$, in the $(\psi,\theta)$ plane and a point, $P_i$, in the Poincare section, the connection length,  $\ell_{B}$, is determined by the number of iterations to go from $P_i$ to ${\cal R}$ and from ${\cal R}$ to $P_i$.  
That is, 
given a point $\left( \psi_i, \theta_i \right)$ in the Poincare section 
$\left \{ \left( \psi_0, \theta_0 \right),  \left( \psi_1, \theta_1 \right), \ldots  \left( \psi_N, \theta_N \right) \right \}$ (where $\left( \psi_k, \theta_k\right)$ for $k=1,\, \ldots N$ denotes the $k$-th crossing starting from the initial condition $\left( \psi_0, \theta_0 \right)$) we determine the smallest integer $n_+$ such that 
$\left( \psi_{i+n_+}, \theta_{i+n_+} \right) \in {\cal R}$ and the smallest integer $n_-$ such that 
$\left( \psi_{i-n_-}, \theta_{i-n_-} \right) \in {\cal R}$ and define
\bq
\label{cl}
\ell_B = \min \{ n_+, n_-\} \, . 
\eq
If $n_+$ exists but $n_-$ does not exist, we define $\ell_B =n_+$.
Conversely, if  $n_-$ exists but $n_+$ does not exist, $\ell_B =n_-$. In the event when neither $n_+$ nor $n_-$ exist, $\ell_B= \infty$.  The region  ${\cal R}$ is defined as 
\bq
\label{region}
{\cal R} = \left \{ \left( \psi, \theta \right) \left | \right. \psi_1 < \psi < \psi_2 \right \}  \, ,
\eq
where $\psi \in (\psi_1, \psi_2)$ is the region where the power modulation heat source is radially localized. 
Based on Eq.~(\ref{source_rt}), we set
$R \psi_1=0.314$ and $R \psi_2=0.336$ which defines an interval of width $R \psi_2-R \psi_1 \sim 3 \sigma$
centered at $R \psi_0=0.325$. 

\begin{figure}
\includegraphics[width=0.75\columnwidth]{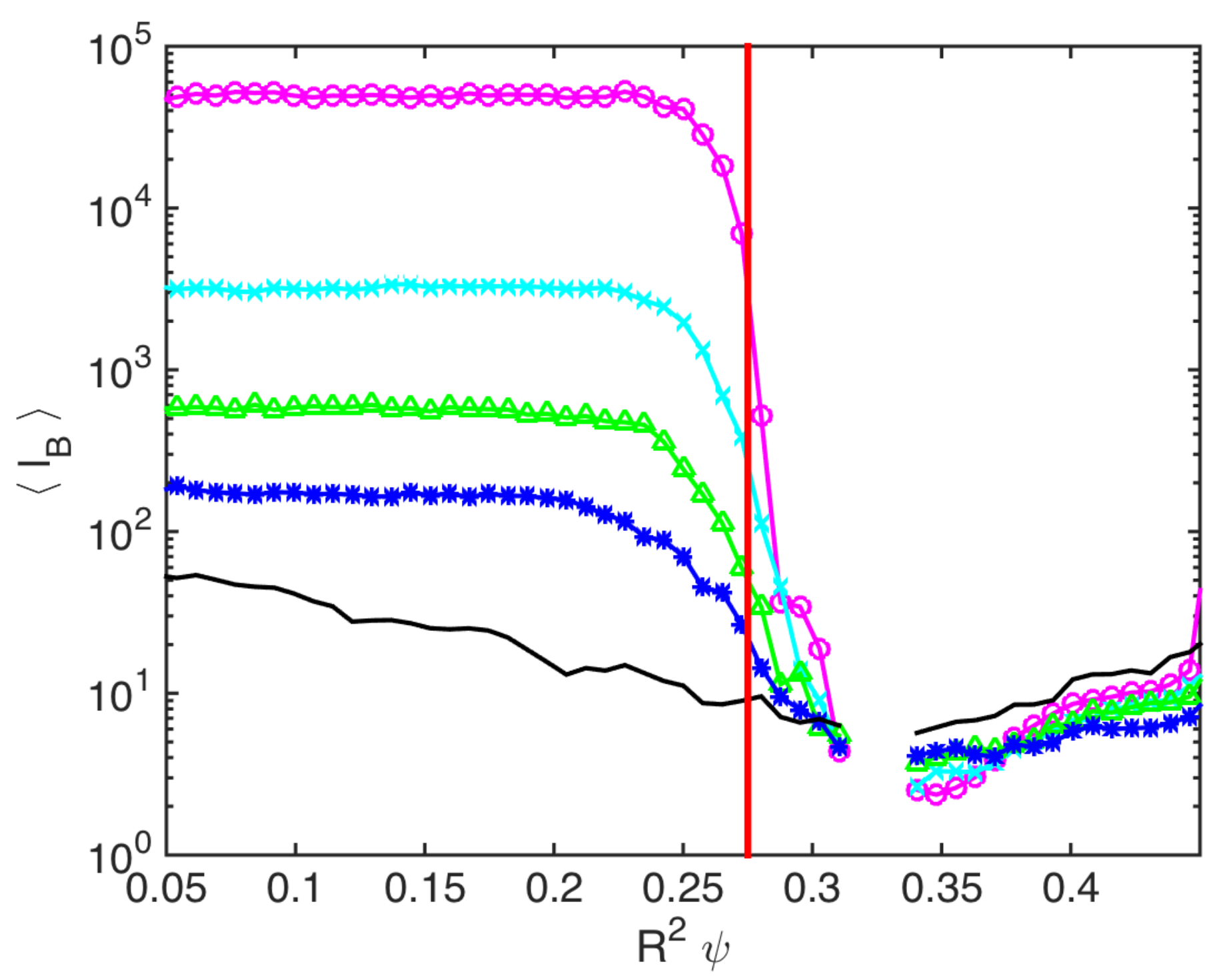}
\caption{
\label{profiles_connection_fig}
(Color online) Magnetic field connection length averaged on $\theta$, $\langle \ell_B \rangle$, as function of 
$R^2 \psi$ for different levels of magnetic field stochasticity: $\epsilon=1.5 \times 10^{-4}$ (Magenta circles); 
$2.0 \times 10^{-4}$ (Cyan crosses); 
$2.5 \times 10^{-4}$ (Green triangles); and $3.0 \times 10^{-4}$ (Blue stars). The solid black line denotes the fully chaotic case.  The interval $\psi \in \left(0.31, 0.34\right)$ indicates the location of the source in Eq.~(\ref{source_rt}), and the vertical red line at $R^2 \psi =0.275$ denotes the location of the partial barrier. [Adapted from Ref.~\cite{diego_dan_2014}]}
\end{figure}

As expected, the connection length increases with the distance from the location of the source and it exhibits a monotonically decreasing dependence on the perturbation amplitude $\epsilon$. Far less trivial, and key for the understanding of transport, is the fractal like distribution of connection lengths and, most importantly, the sudden increase by several orders of magnitude of the connection length in the region around $R^2 \psi \approx 0.275$ (marked in Figs.~\ref{poinc_connection_fig} and 
\ref{profiles_connection_fig} with a vertical red line). The increase in the connection length, that manifests as a strong gradient in the radial profiles shown in Fig.~\ref{profiles_connection_fig}, is caused by the rich structure of high order islands and the presence of  geometric structures known as Cantori. 

%%%%%%%%%%%%%%%%%%%%%%%%%%%%%%%%%%%%
%%%%%%%%%%%%%%%%%%%%%%%%%%%%%%%%%%%%

The existence of Cantori in Hamiltonian systems was originally proposed in Ref.~\cite{percival}  and their impact on transport first studied in Ref.~\cite{mckay}. More recently, the role of Cantori on anisotropic heat transport in plasmas has been studied in the context of steady state solutions \cite{hudson} and time dependent solutions in monotonic and reversed shear magnetic configurations \cite{dan_diego_2013,diego_dan_2014}.
The basic idea is that although invariant closed curves (which correspond to flux surfaces in the Hamiltonian description of magnetic fields) present impenetrable transport barriers, regions without such curves (e.g., the region around 
$R^2 \psi \approx 0.275$ in Fig.~\ref{poinc_connection_fig}) contain invariant Cantor sets.
These Cantor sets, known as Cantori, are characterize by an infinite number of gaps through which orbits can ``leak". However, since these structures have finite length, the gaps are very small and transport across them  is remarkably slow.  As a result,  Cantori act as partial barriers. For example, according to Fig.~\ref{profiles_connection_fig},  for $\epsilon = 1.5 \times 10^{-4}$ it takes on the average $5 \times 10^4$ iterations for a magnetic field initial located in the source region, 
$R^2 \psi \approx 0.336$,  to reach the region 
$R^2 \psi \approx 0.2$, but it takes only about $10$ iterations to reach the region $R^2 \psi \approx 0.45$. 

\begin{figure}
\includegraphics[width=0.50\columnwidth]{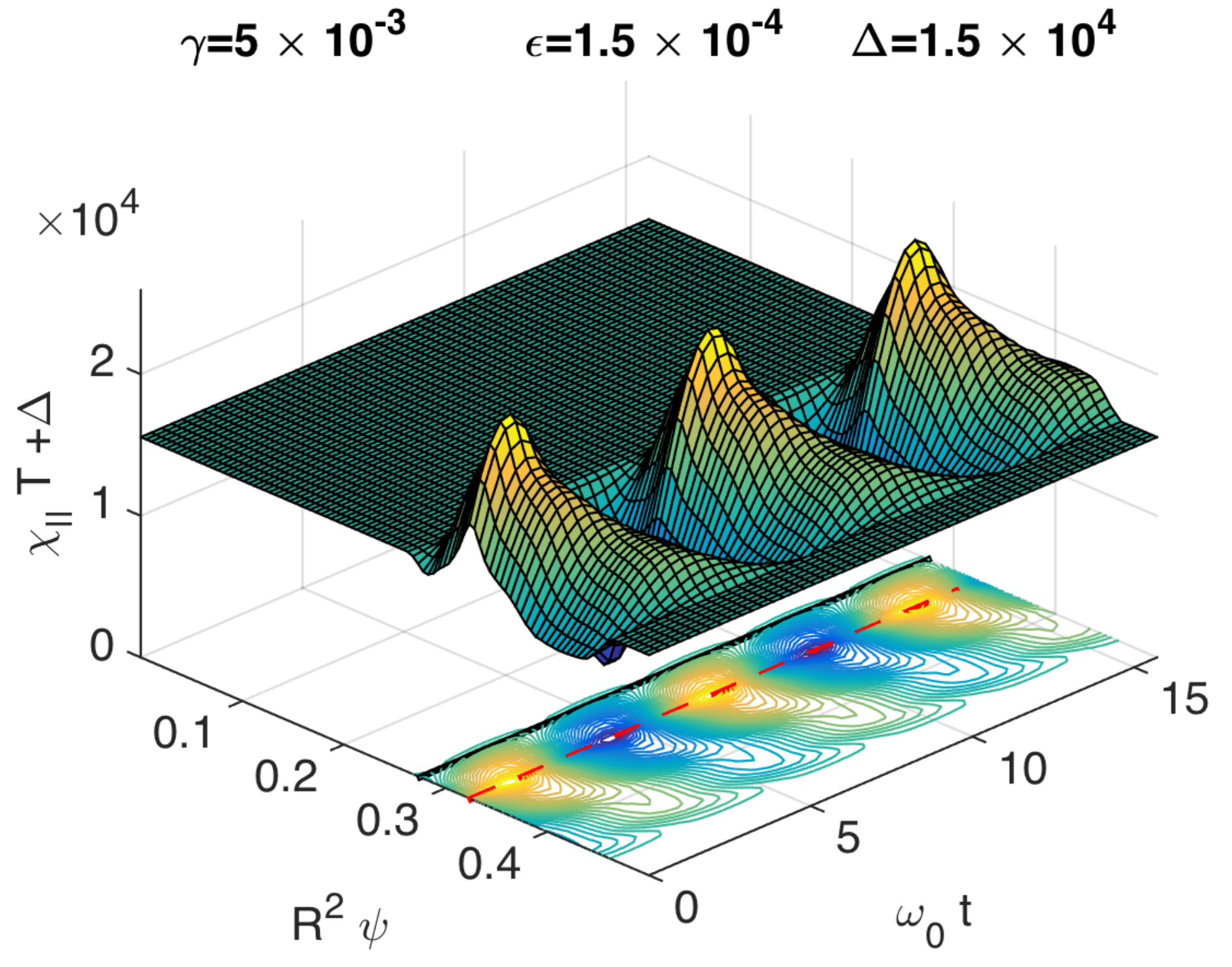}
\includegraphics[width=0.50\columnwidth]{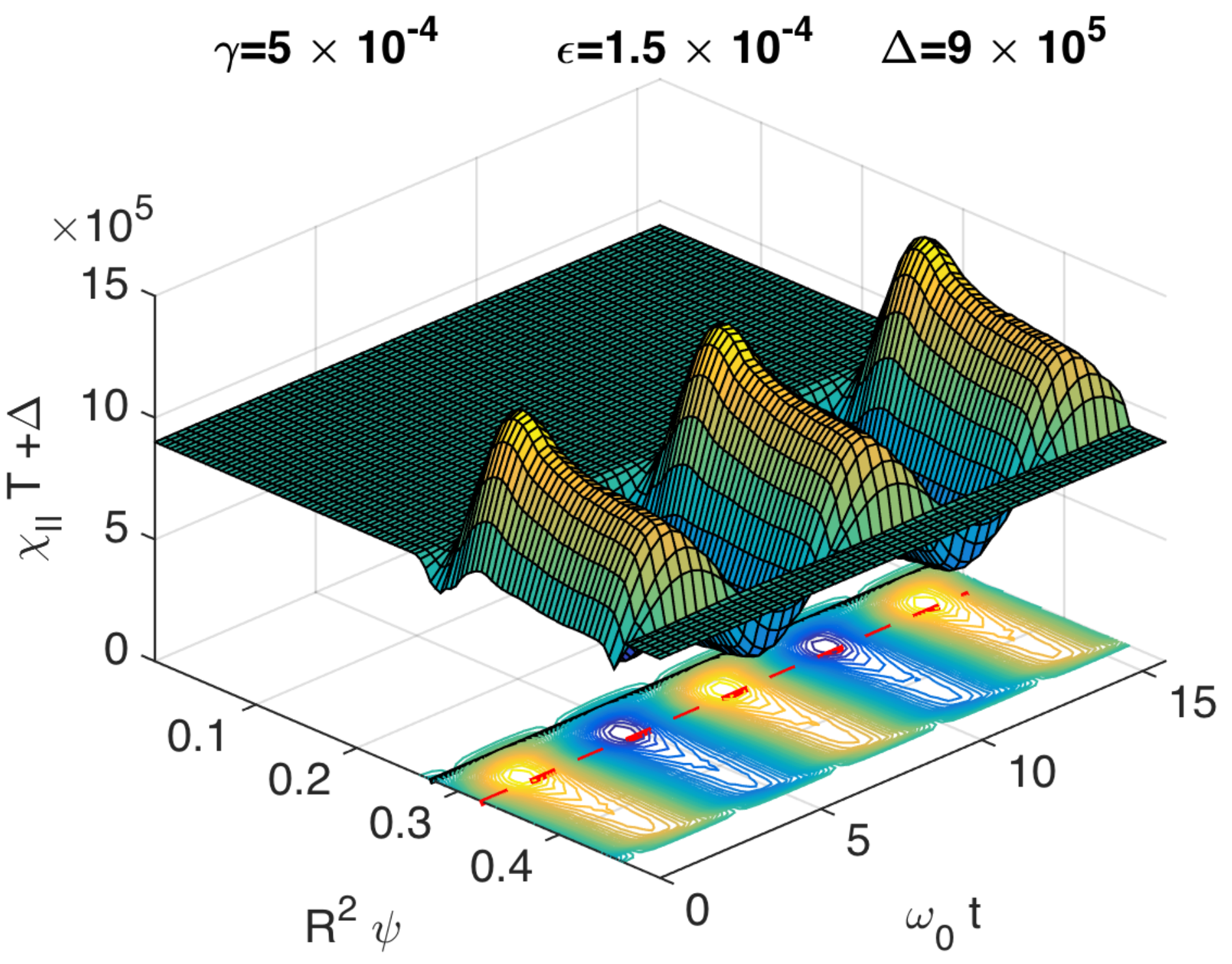}
\includegraphics[width=0.50\columnwidth]{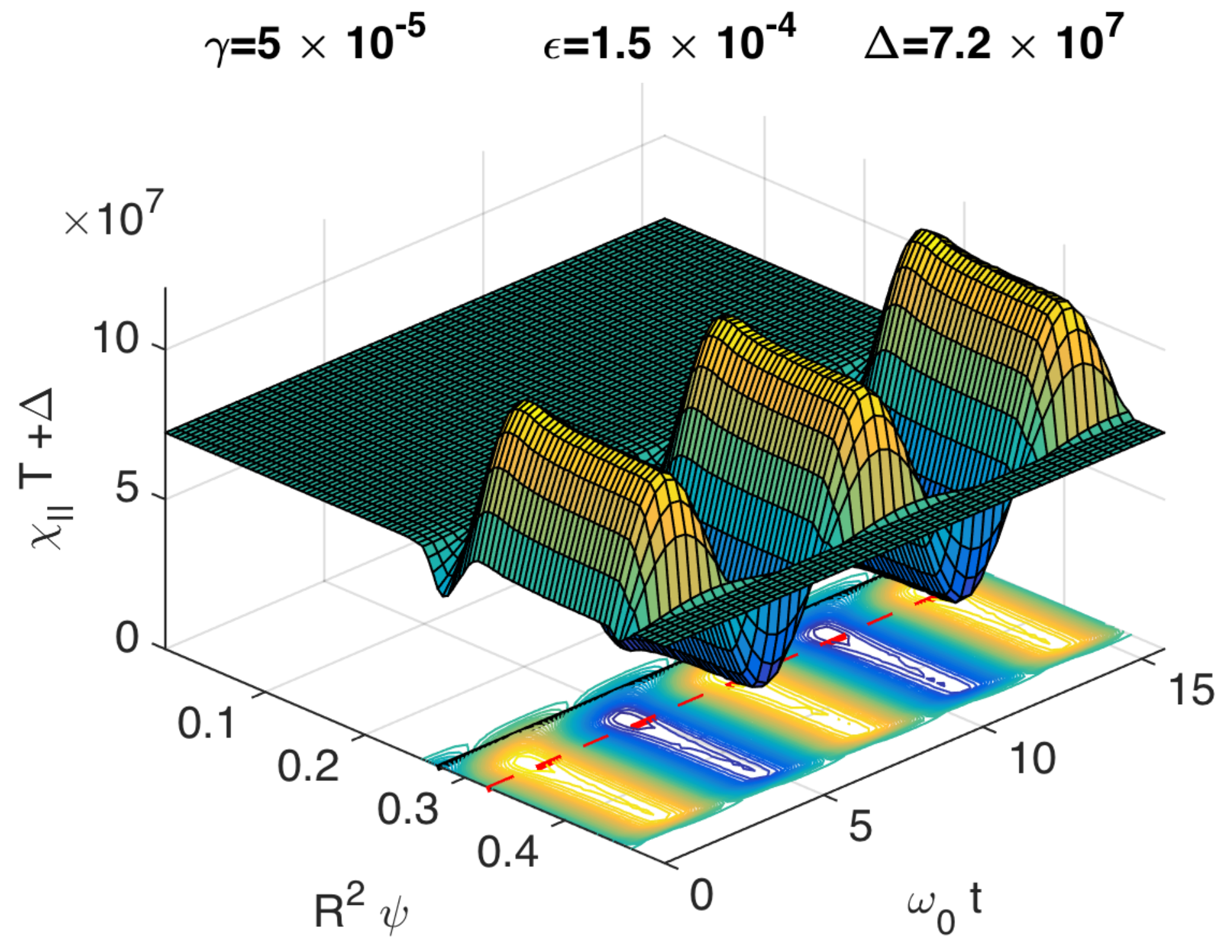}
\caption{(Color online) 
Space-time evolution of temperature modulation amplitude
(averaged over $z$ and $\theta$) for weak ($\epsilon=1.5 \times 10^{-4}$) stochasticity and different values of inverse 
penetration length scale $\gamma$ in Eq.~(\ref{gamma_def}). For illustration purposes the vertical axis has been shifted by $\Delta$. The dashed red line denotes the location of the source in Eq.~(\ref{source_rt}), and the solid black line at $R^2 \psi =0.275$ denotes the location of the partial barrier.}
\label{T_space_time_weak_stocha}
\end{figure}

\begin{figure}
\includegraphics[width=0.50\columnwidth]{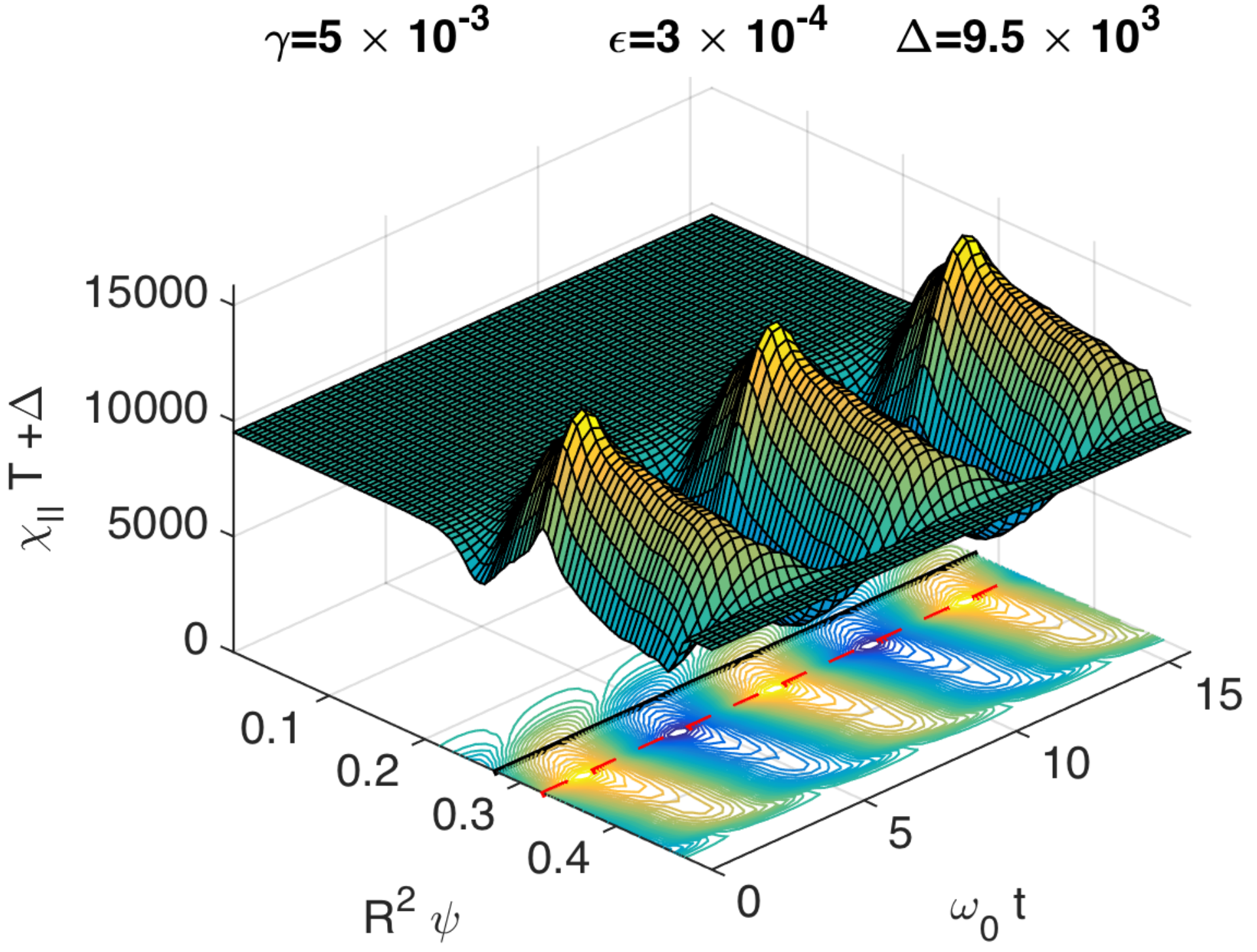}
\includegraphics[width=0.50\columnwidth]{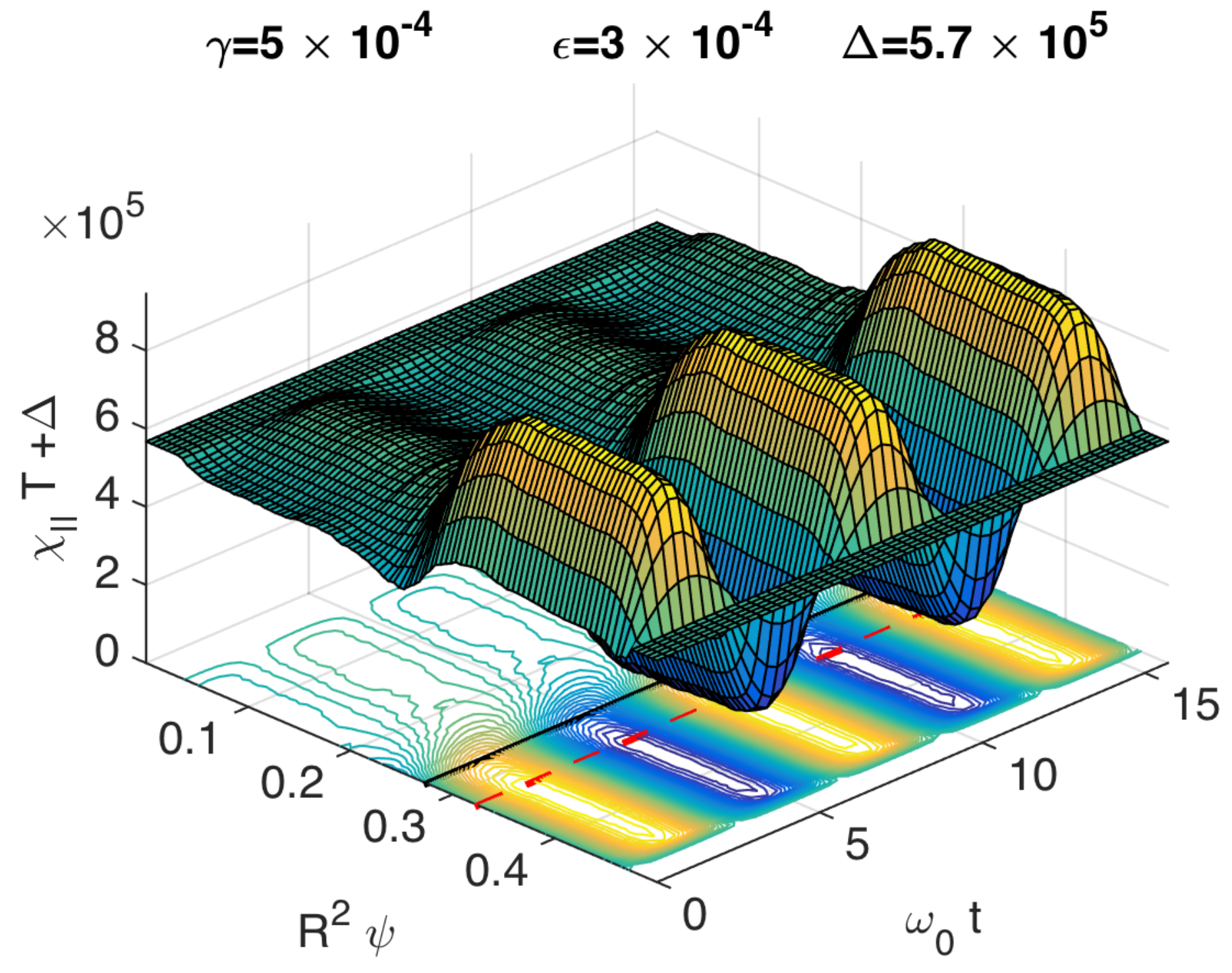}
\includegraphics[width=0.50\columnwidth]{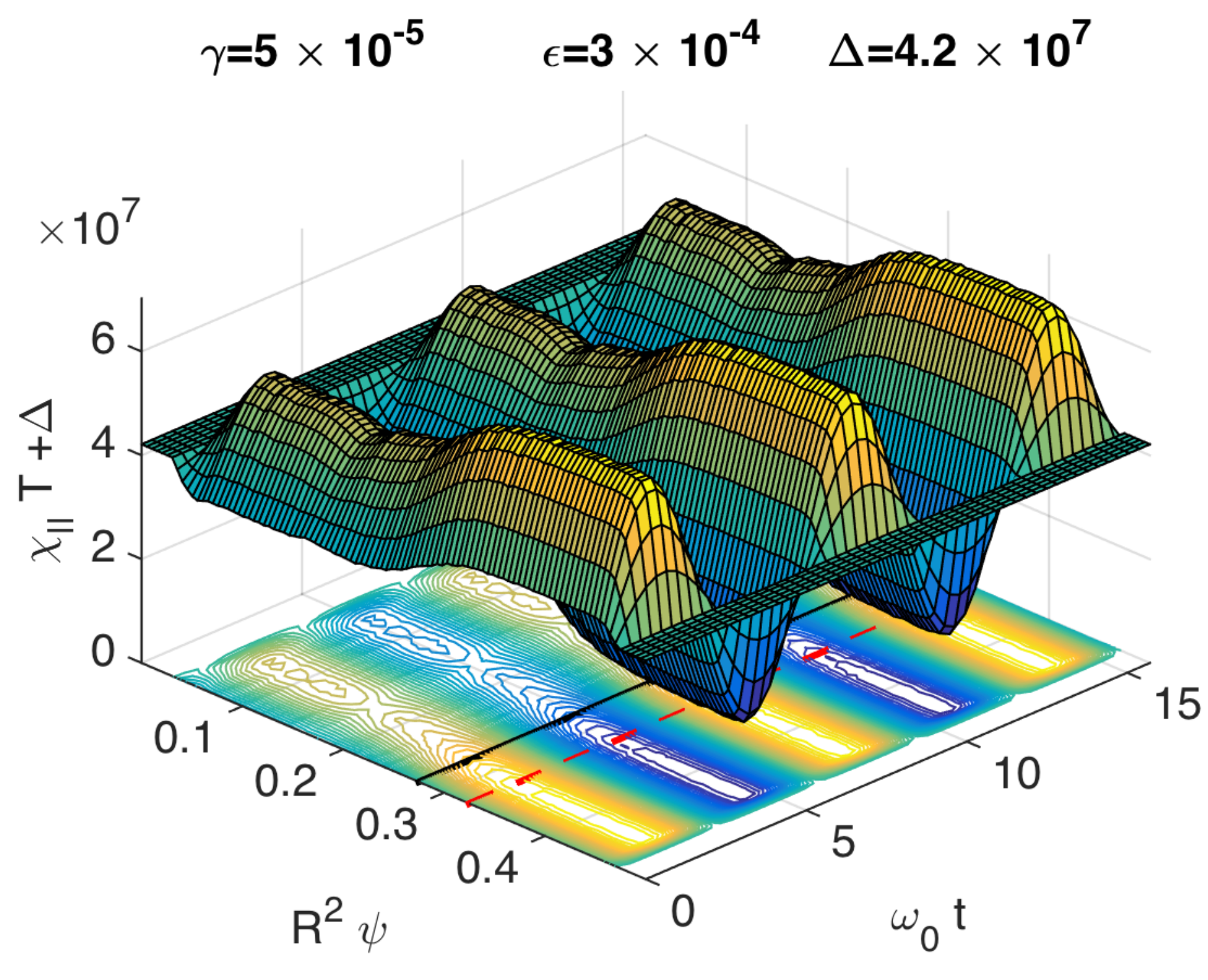}
\caption{(Color online) Same as Fig.~\ref{T_space_time_weak_stocha} but for 
moderate ($\epsilon=3.0 \times 10^{-4}$) stochasticity.
}
\label{T_space_time_moderate_stocha}
\end{figure}

\begin{figure}
\includegraphics[width=0.4\columnwidth]{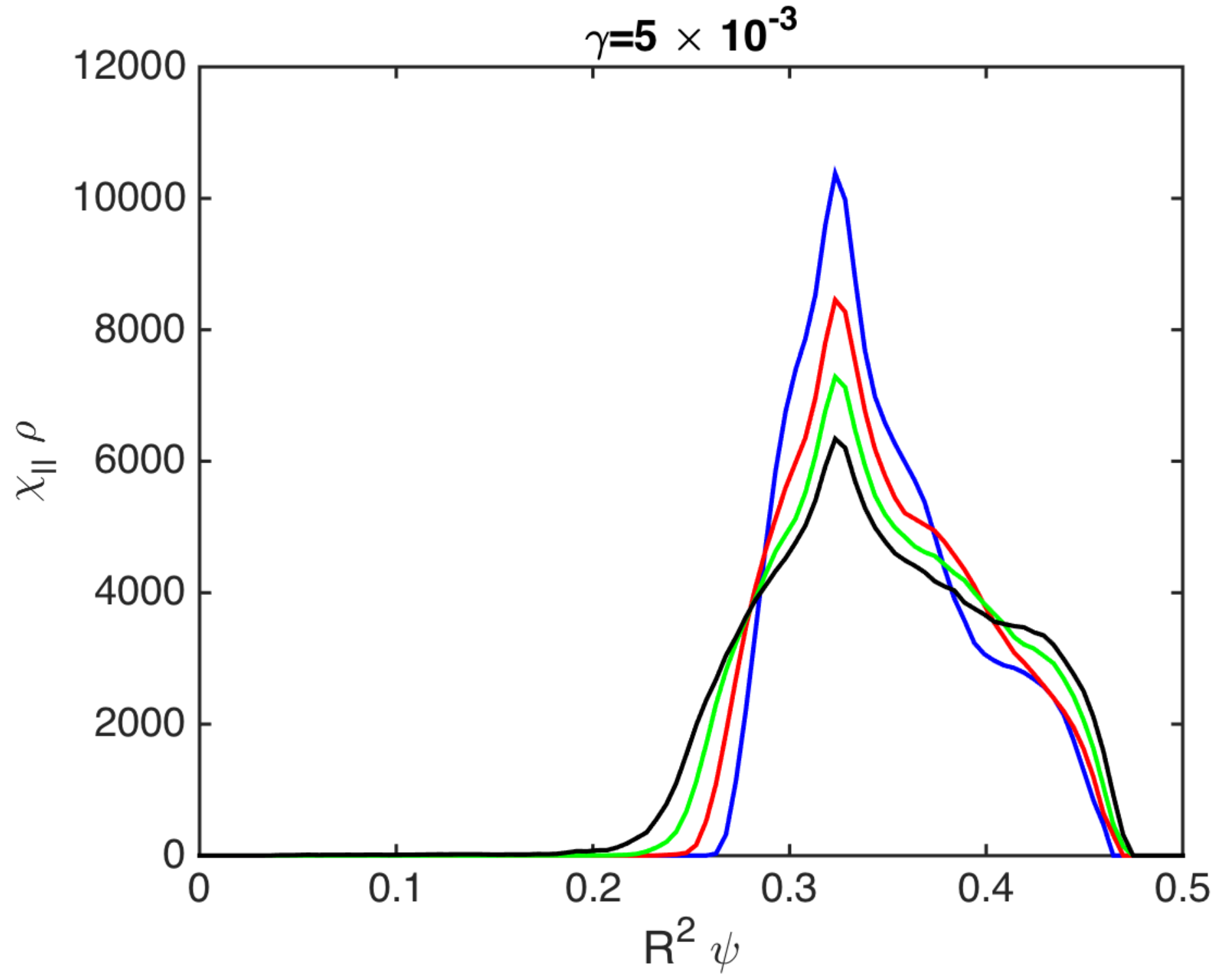}
\includegraphics[width=0.4\columnwidth]{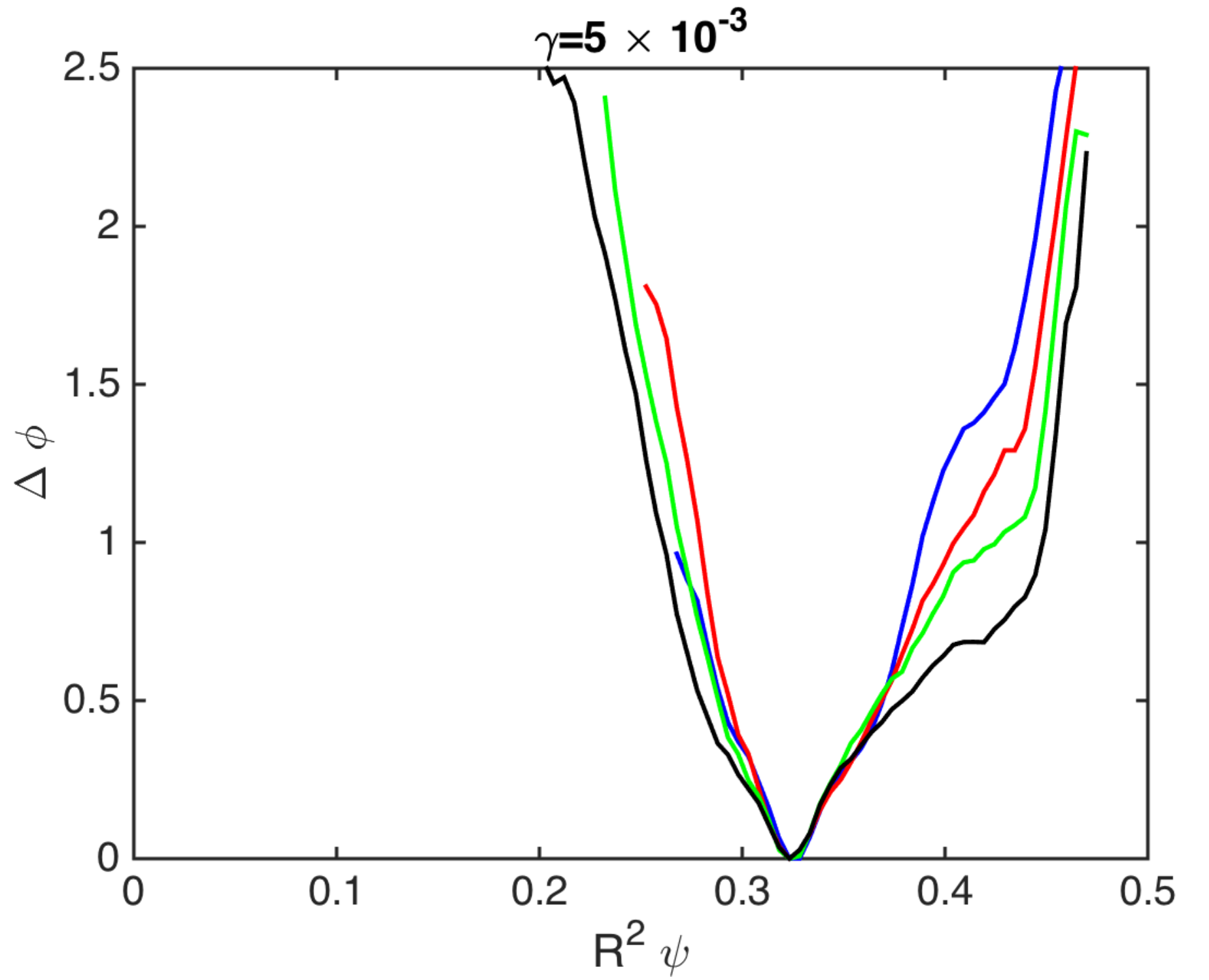}
\includegraphics[width=0.4\columnwidth]{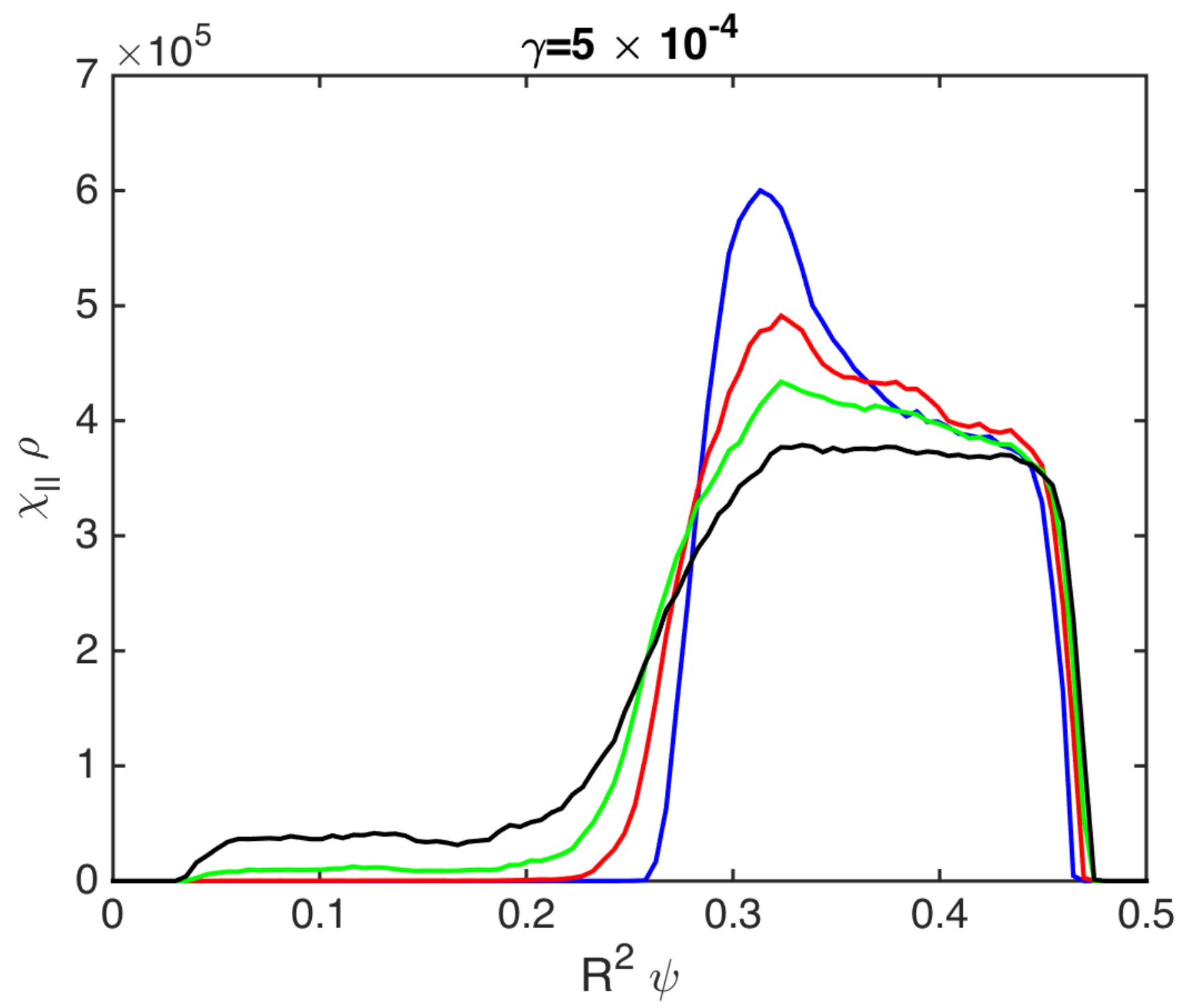}
\includegraphics[width=0.4\columnwidth]{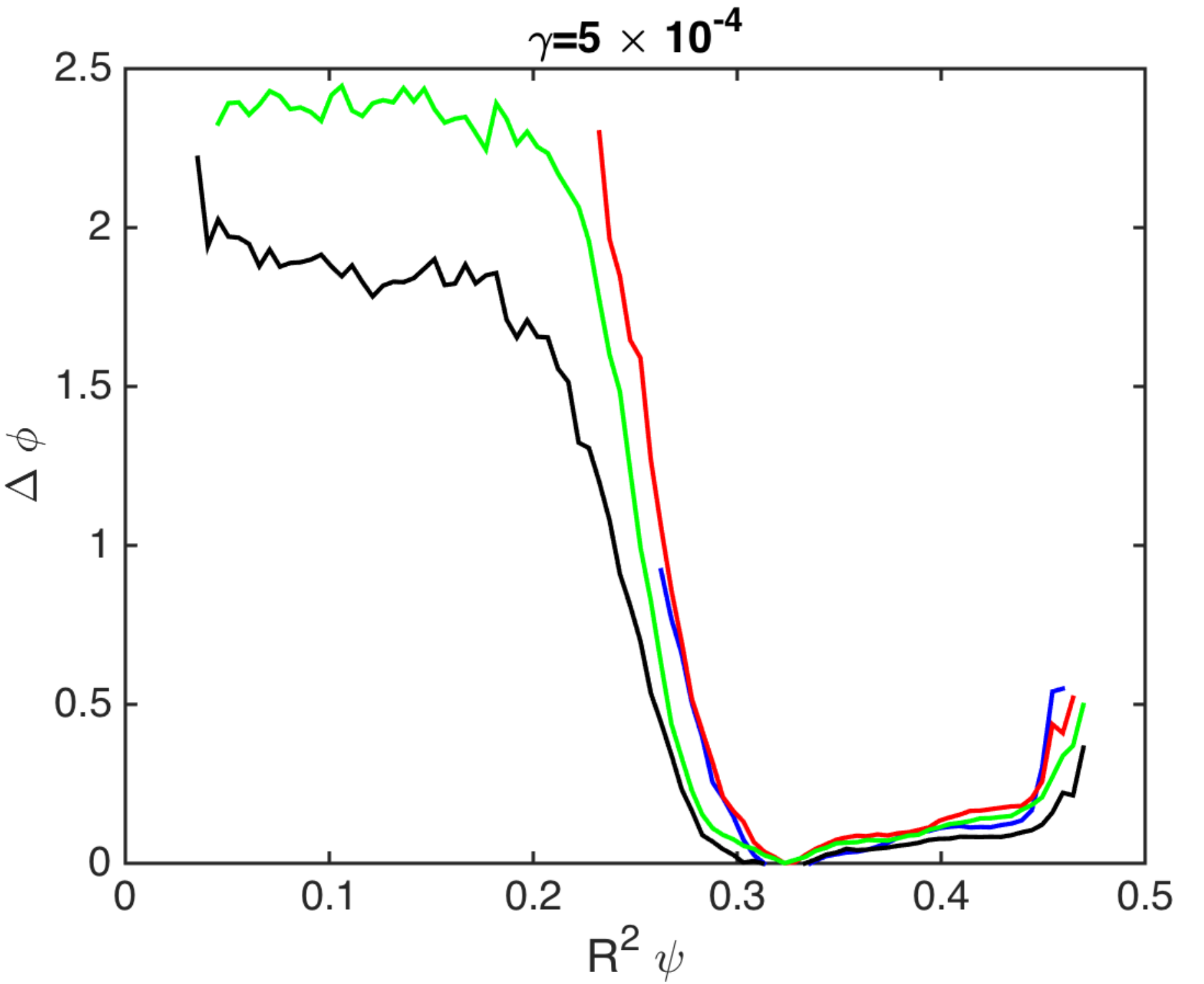}
\includegraphics[width=0.4\columnwidth]{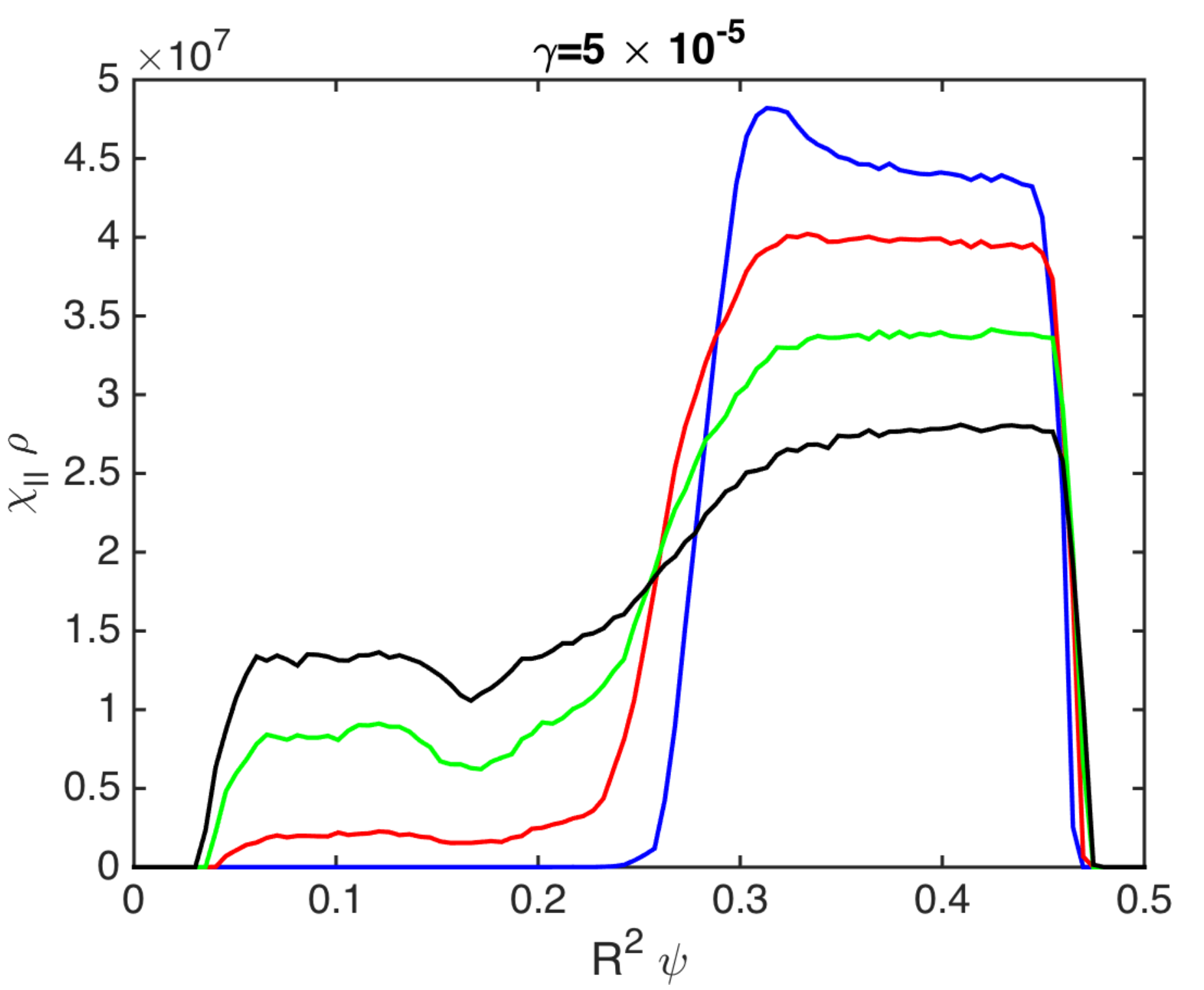}
\includegraphics[width=0.4\columnwidth]{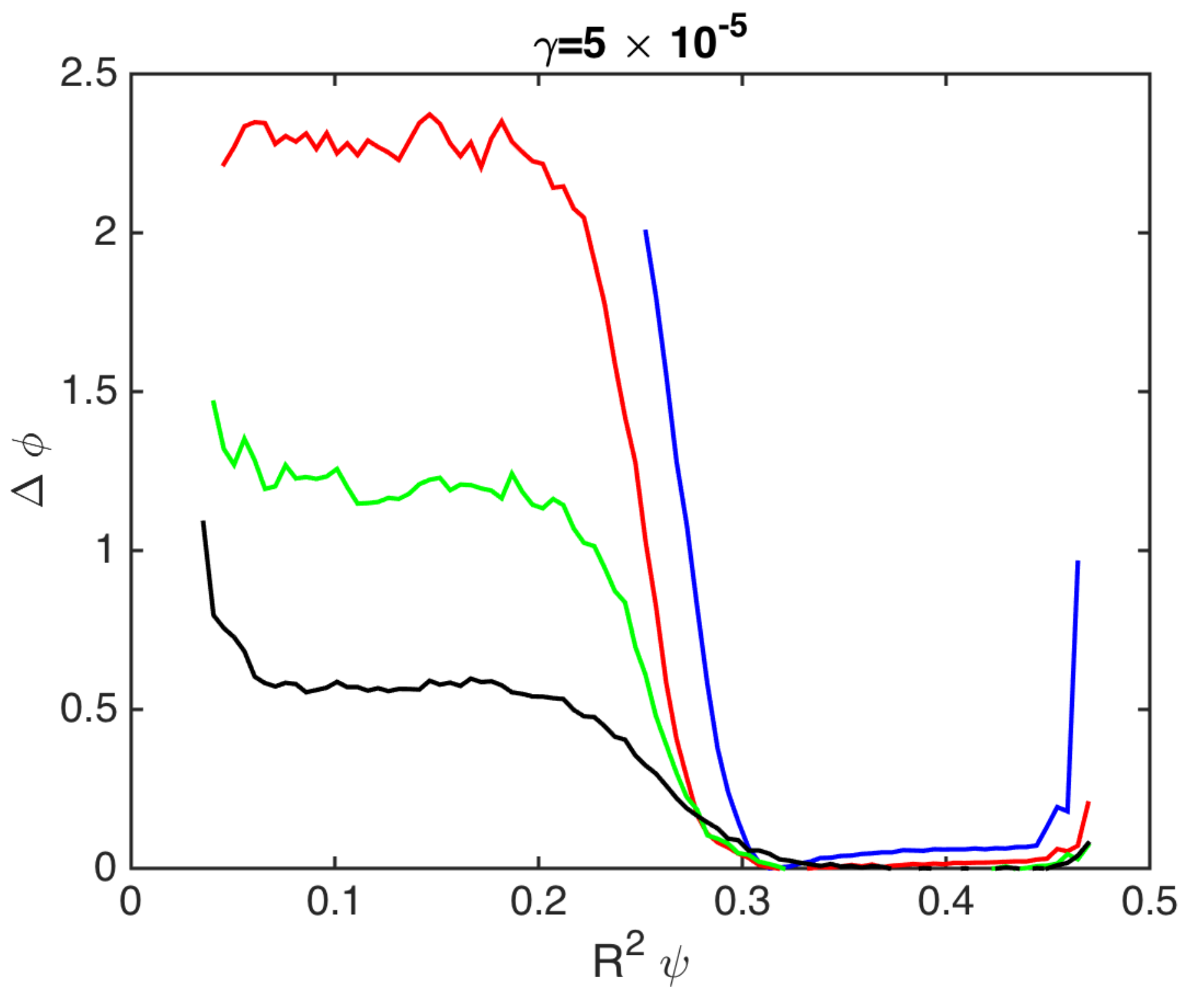}
\caption{(Color online) Temperature modulation amplitude, $\rho$, and relative phase, $\Delta \phi$,  for different values of 
penetration inverse length scale $\gamma$. In each figure the different curves correspond to different values of the stochasticity parameter $\epsilon$. Blue $\epsilon=1.5 \times 10^{-4}$; Red $\epsilon=2.0 \times 10^{-4}$; Green $\epsilon=2.5 \times 10^{-4}$; Black $\epsilon=3.0 \times 10^{-4}$.}
\label{T_profiles_weak_moderate_ch}
\end{figure}

\begin{figure}
\includegraphics[width=0.45\columnwidth]{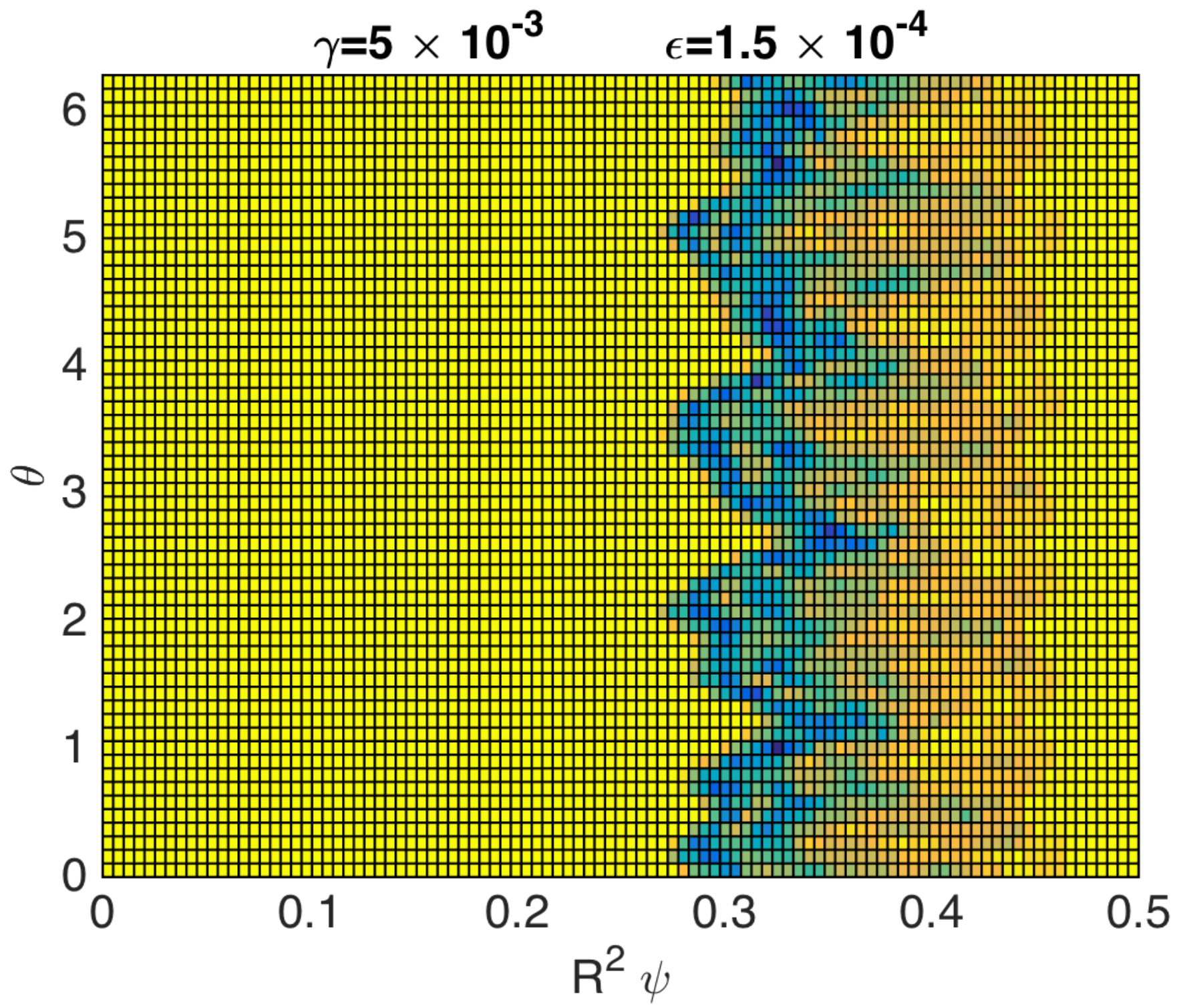}
\includegraphics[width=0.45\columnwidth]{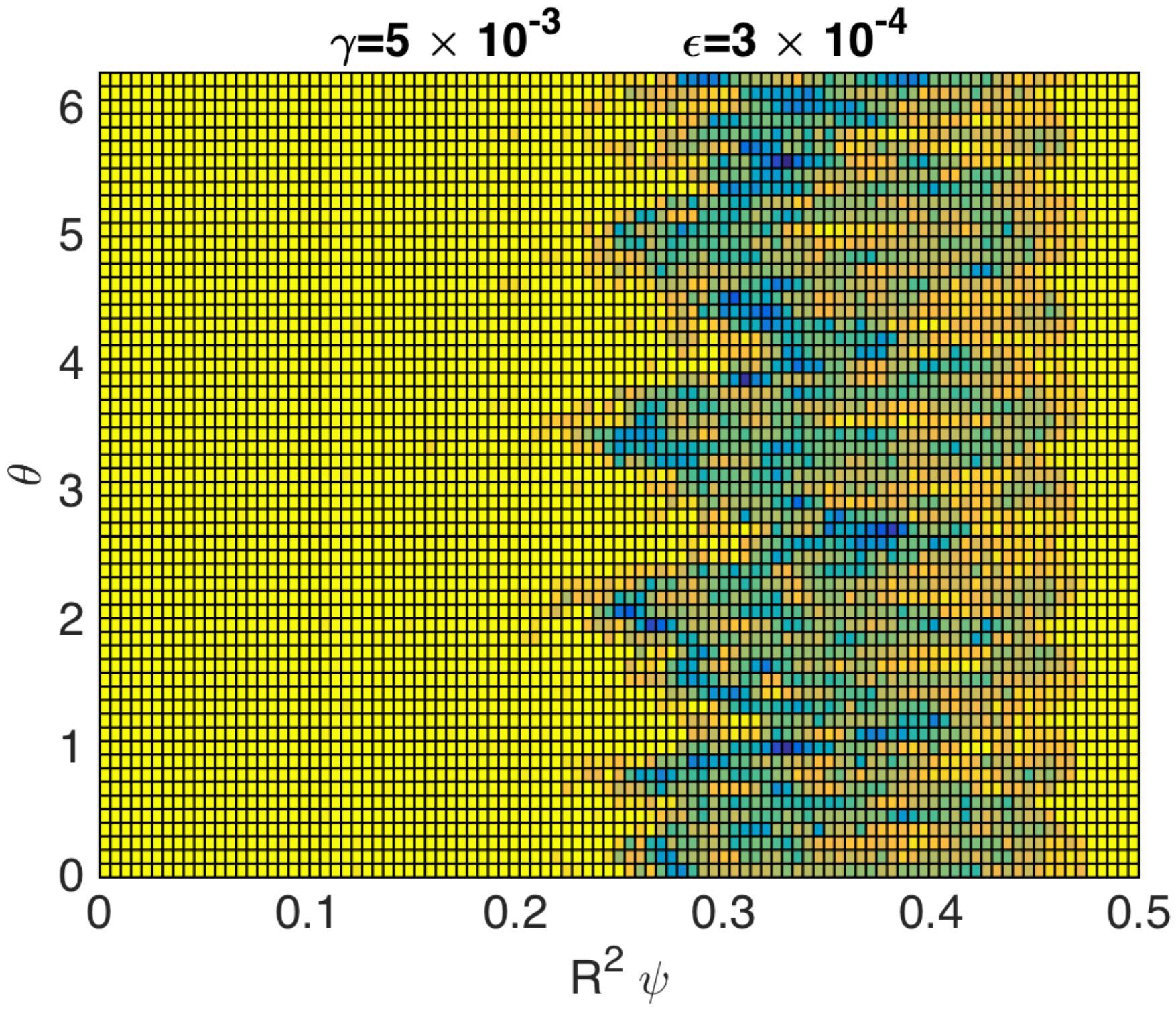}
\includegraphics[width=0.45\columnwidth]{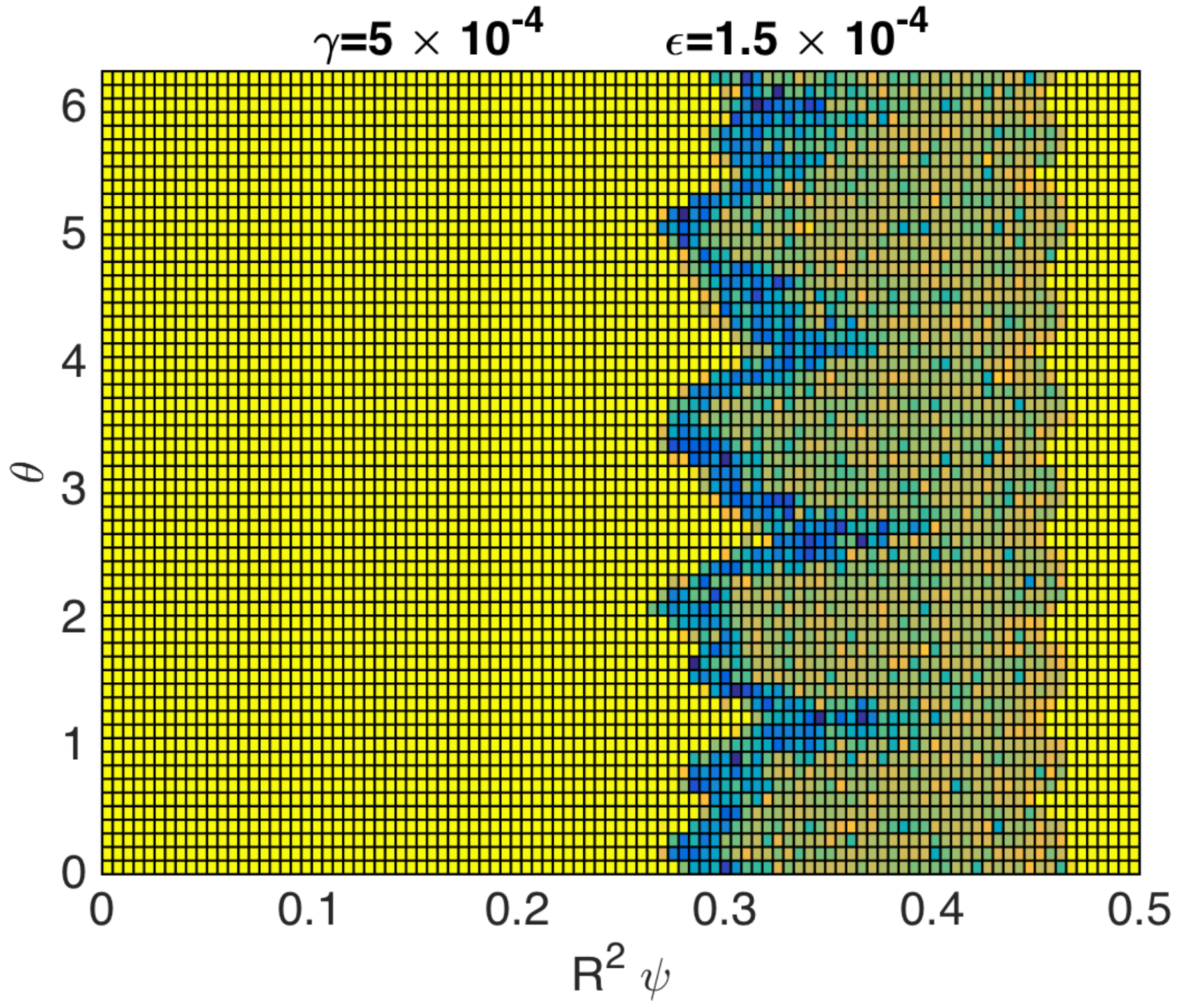}
\includegraphics[width=0.45\columnwidth]{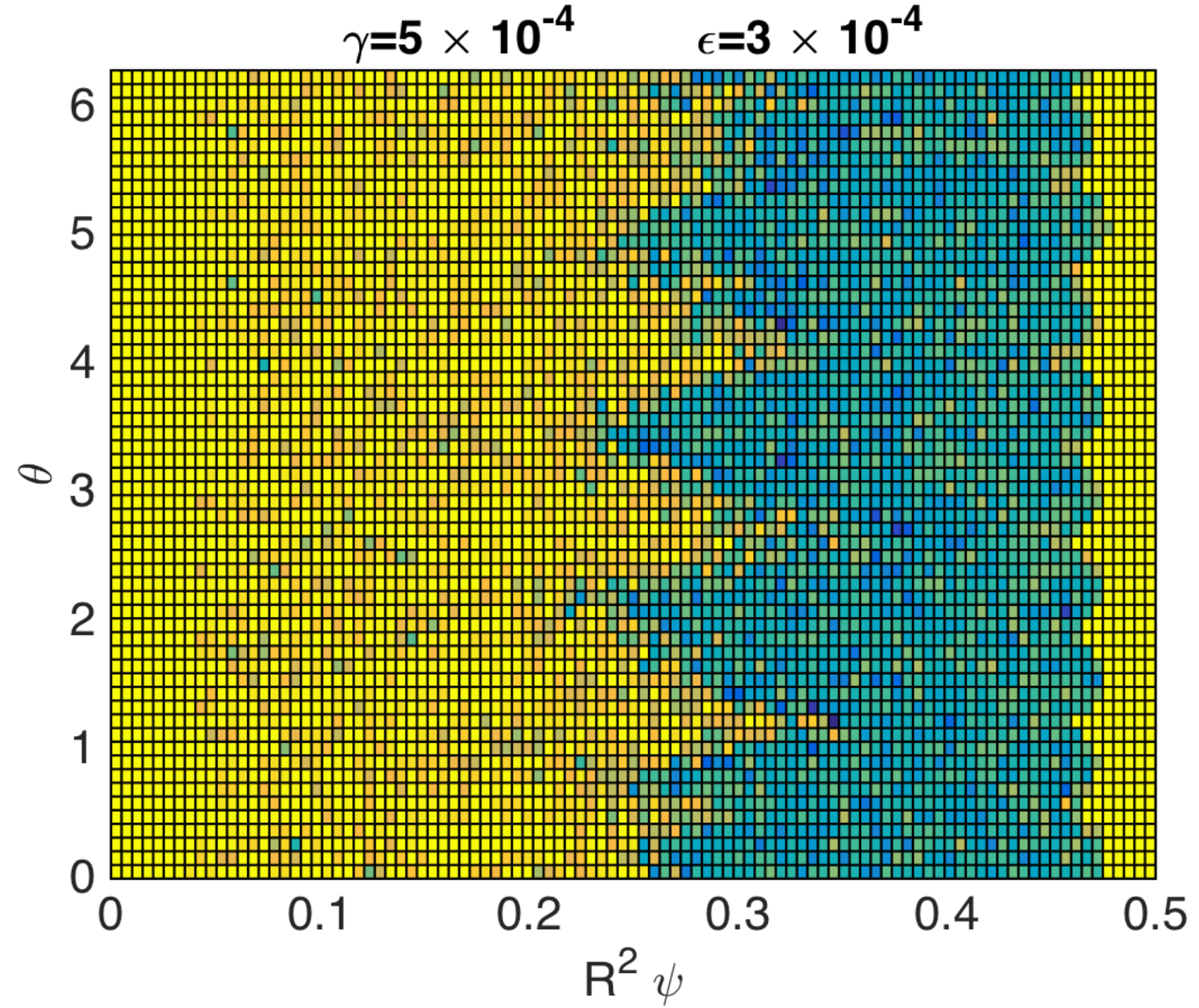}
\includegraphics[width=0.45\columnwidth]{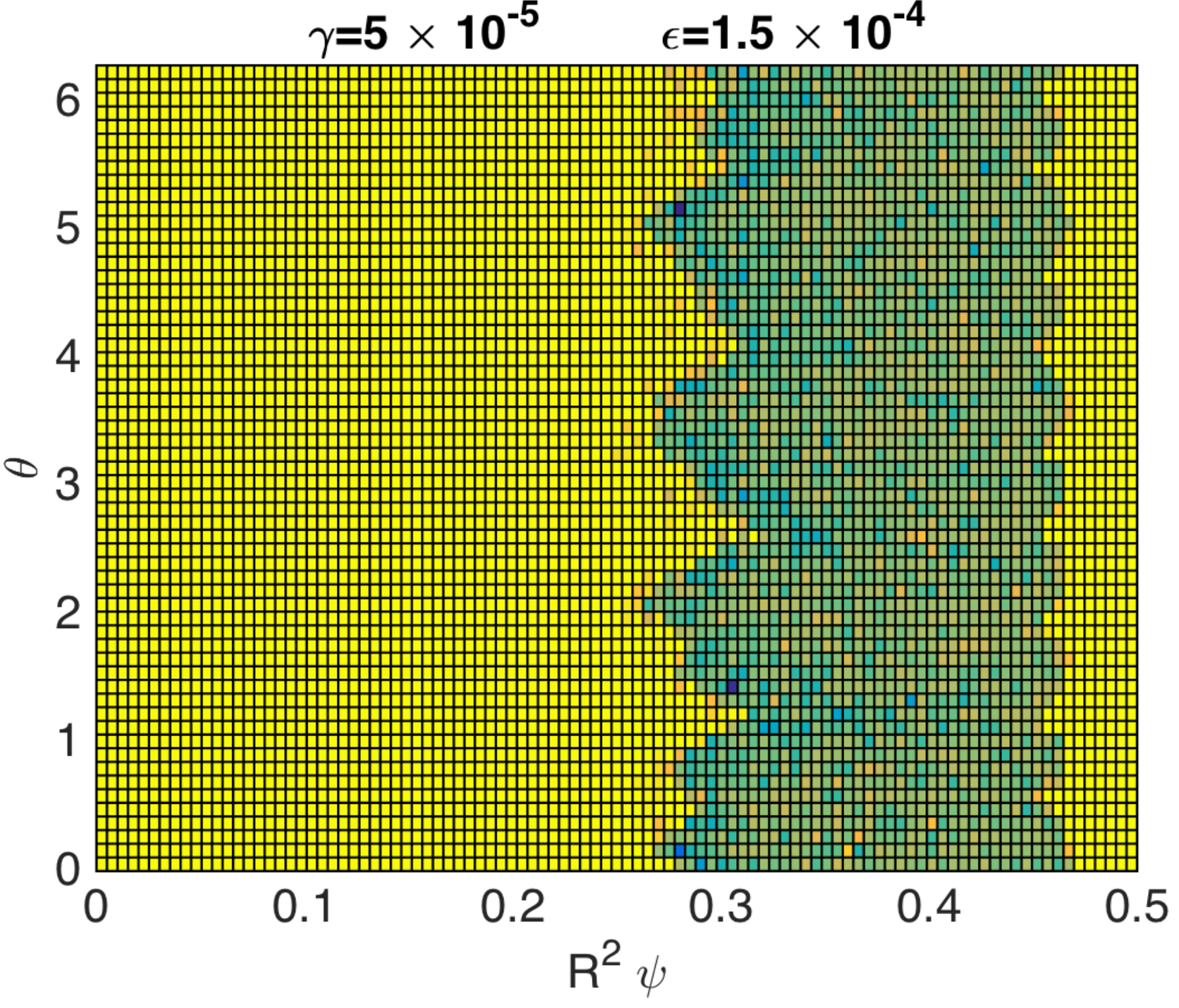}
\includegraphics[width=0.45\columnwidth]{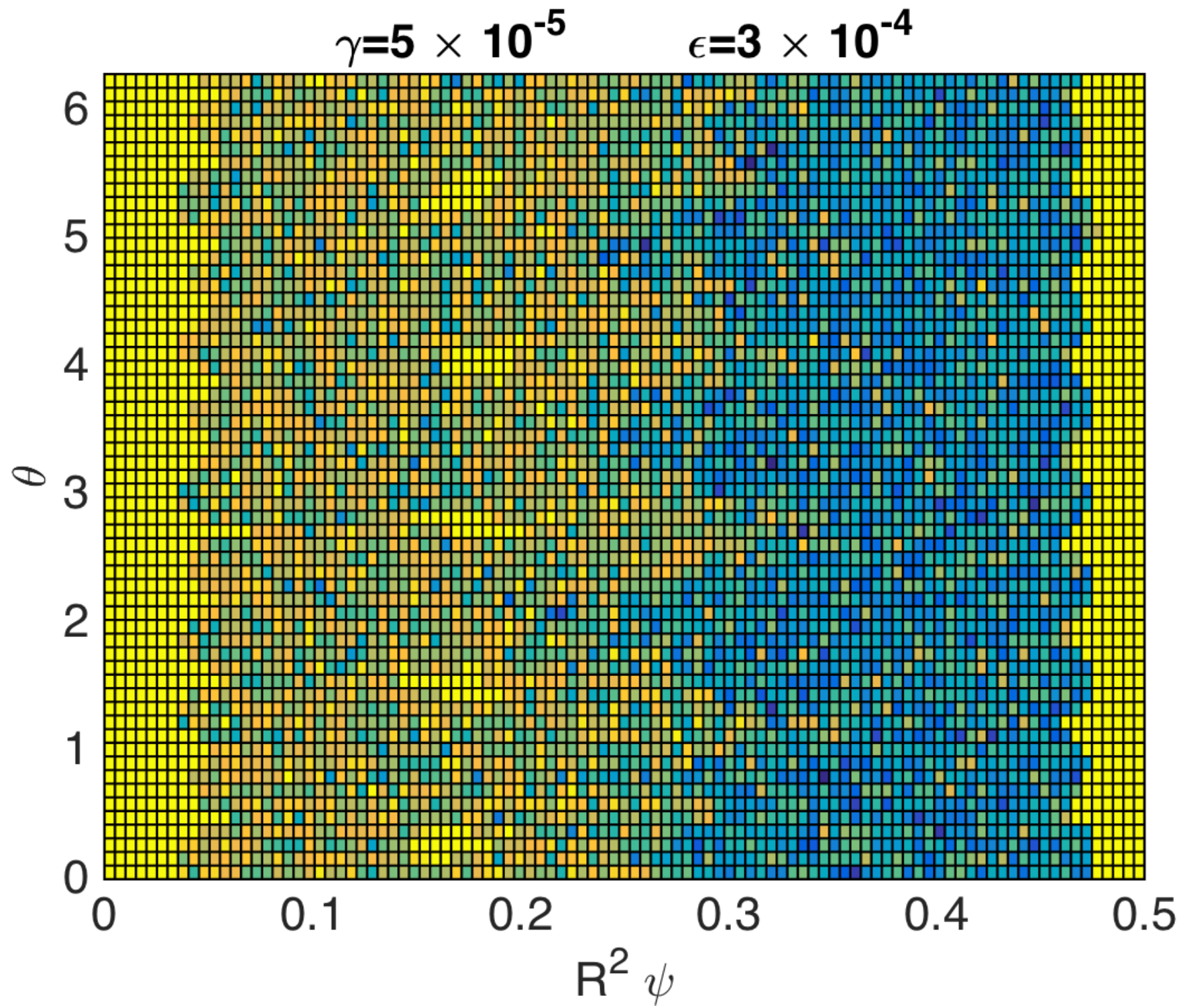}
\caption{(Color online) 
Spatial dependence in $(R^2 \psi, \theta)$ plane (at fixed $z$)  of temperature modulation amplitude
in Eq.~(\ref{rho_phi_def}) for weak ($\epsilon=1.5 \times 10^{-4}$) and moderated ($\epsilon=3 \times 10^{-4}$) stochasticity and different values of inverse 
penetration length scale $\gamma$ in Eq.~(\ref{gamma_def}). Dark blue (light yellow) corresponds to large (small) values. 
For the quantitative ranges of amplitude values refer to Fig.~\ref{T_profiles_weak_moderate_ch}.}
\label{T_psi_theta_weak_moderate_ch}
\end{figure}

Figures~\ref{T_space_time_weak_stocha}, \ref{T_space_time_moderate_stocha}, \ref{T_profiles_weak_moderate_ch}, and \ref{T_psi_theta_weak_moderate_ch} summarize the numerical results for the propagation of modulated heat pulses in the magnetic field perturbed by the four modes mentioned above according to the Lagrangian-Green's function solution in Eq.(\ref{LG-F}) with $m=1$. The plots in Figs.~\ref{T_space_time_weak_stocha} and \ref{T_space_time_moderate_stocha} show the space-time evolution of the temperature averaged over $z$ and $\theta$,
$\langle T \rangle (\psi, t) =(1/2\pi L) \int_0^L dz \int_0^{2 \pi} d\theta \, T({\bf r},t)$, where $L=2 \pi R$ is the length of the periodic cylindrical domain, and the plots in Fig.~\ref{T_profiles_weak_moderate_ch} show the radial dependence of the wave amplitude $\rho=\rho(R^2 \psi)$ and the phase $\Delta \phi=\Delta \phi(R^2 \psi)$ where
\bq
\label{rho_phi_def}
\langle T({\bf r}_0,t) \rangle= {\rm Re} \frac{\langle  {\cal I}_1\rangle }{ \chi_\parallel \left(1 + i\right) \gamma } e^{i \omega_0 t} =
{\rm Re} \,\, \chi_\parallel \rho  \, e^{i \left[\omega_0 t+ \Delta \phi  \right]} \, .
\eq
We consider two cases with perturbation amplitudes, $\epsilon=1.5 \times 10^{-4}$ and 
$\epsilon=3.0 \times 10^{-4}$, and for each case we consider three values of the inverse penetration length scale defined  in Eq.~(\ref{gamma_def}), namely $\gamma=5 \times 10^{-3}$, $\gamma=5 \times 10^{-4}$ and $\gamma=5 \times 10^{-5}$.
As expected, the temperature peaks at the location of the source $R^2 \psi \approx 0.336$ and decays away from it. However, the rate of decay is remarkably different  for $R^2 \psi <  0.336$ and $R^2 \psi >  0.336$. In particular, due to the existence of partial barriers in the region $R^2 \psi \approx 0.275$,  the heat wave is strongly damped for 
$R^2 \psi < 0.275$. The results depend on both the strength of the perturbation, $\epsilon$, and the inverse penetration length $\gamma$. In general,  the damping of $\rho$ is proportional $\gamma$ and inverse proportional to $\epsilon$. The critical dependence on $\gamma$ indicates that care should be taken when inferring properties of the magnetic filed topology using power modulation perturbative experiments with a {\em fixed} frequency. For example, what might be interpreted as a transport barrier when  $\epsilon=3.0 \times 10^{-4}$ based on a high-frequency (e.g., $\gamma=5 \times 10^{-3}$) power modulation might be interpreted as the lack of a transport barrier for the {\em same} magnetic field when when using a low-frequency  perturbation (e.g., $\gamma=5 \times 10^{-5}$). In addition to the heat wave amplitude, $\rho$, the radial dependence of the phase shift, $\Delta \phi$,  provides valuable transport information. By definition,
at the location of the source, $\Delta \phi(R^2 \psi=0.325)=0$ and away from the source  $\Delta \phi>0$ due to the 
finite propagation speed of the heat wave. In particular, $\tau(\psi)=\Delta \phi /\omega_0$ gives the time delay of the perturbation as a function of $\psi$, and the inverse of the local slope of the $\Delta \phi$ v.s. $\psi$ plot gives an estimate of the local phase speed of the heat wave, $V\sim d \psi / d \tau$. 
Note that the damping of the heat wave while crossing the partial transport barrier is accompanied by a significant slowing down of the heat wave speed around $R^2 \psi \approx 0.275$ and the corresponding increase of the time delay. 
In the cases when the perturbation manages to go through the partial barrier maintaining a non vanishing finite amplitude 
(e.g., when $\gamma=5\times 10^{-5}$ and $\epsilon > 2.0 \times 10^{-4}$) the local phase speed of the heat wave
increases significantly in response to the high level of stochasticity in that region (see Fig.~\ref{poinc_connection_fig}) and the negligible parallel damping (which according to Eq.~(\ref{LG-F}) is proportional to $\gamma$) of the perturbation along the magnetic field. 

%%%%%%%%%%%%%%%%%%%%%%%%%%%%%%%%%%%%
%%%%%%%%%%%%%%%%%%%%%%%%%%%%%%%%%%%%

%%%%%%%%%%%%%%%%%%%%%%%%%%%%%%%%%%%%
%%%%%%%%%%%%%%%%%%%%%%%%%%%%%%%%%%%%

%%%%%%%%%%%%%%%%%%%%%%%%%%%%%%%%%%%%
%%%%%%%%%%%%%%%%%%%%%%%%%%%%%%%%%%%%

The $2$-dimensional dependence of the temperature on the radius and the poloidal angle, $\theta$, at a fixed time is shown in Fig.~\ref{T_psi_theta_strong_ch} for the same values of $\gamma$ and $\epsilon$ discussed above.  
In addition to the already noted radial sharp decay of the amplitude at $R^2 \psi \approx 0.275$  for small values of $\epsilon$ and large values of $\gamma$, it is interesting to observe that the angular distribution of the heat wave amplitude follows a pattern  qualitatively similar to that of the spatial distribution of the magnetic field connection length in Fig.~\ref{poinc_connection_fig} providing further evidence of the key role played by the partial barriers. 

%%%%%%%%%%%%%%%%%%%%%%%%%%%%%%%%%%%%
%%%%%%%%%%%%%%%%%%%%%%%%%%%%%%%%%%%%

%%%%%%%%%%%%%%%%%%%%%%%%%%%%%%%%%%%%%%%%
\subsection{Fully chaotic magnetic fields}
\label{chaos_results}
%%%%%%%%%%%%%%%%%%%%%%%%%%%%%%%%%%%%%%%%

To study the propagation of modulated heat pulses in the presence of fully stochastic fields we consider a perturbation consisting of twenty overlapping modes 
$ (m,n) = \{ (5,4), (9,7), (4,3),(7,5), (3,2),
 (5,3), (7,4),(11,6), (2,1), (9,4), (7,3),
(5,2), (14,5), (3,1),$ $ (10,3), 7,2), (11,3),
( (4,1), (13,3), (9,2), (5,1) \}
$
with amplitude $\epsilon=10^{-4}$.
The values of $m$ and $n$ were chosen so that the resonances are approximately uniformly distributed in the radial $\psi$-domain. To maximize the level of stochasticty  the modes are decorrelated  by choosing the phases of the perturbations, 
$\{ \zeta_{mn}\}$, in Eq.~(\ref{perturbation})  from a random distribution in the interval $[0, 2 \pi]$. 
According to the Chirikov criterion, the strong overlap of the resonances guarantees complete stochasticity, i.e. lack of magnetic surfaces and islands. This is the same magnetic field used in Ref.~\cite{diego_dan_2014} to study freely decaying (i.e., without power modulation) heat pulses in fully stochastic fields. 
As expected, in this case the  magnetic field connection length in Fig.~\ref{profiles_connection_fig} shows a comparatively flat radial profile. 

\begin{figure}
\includegraphics[width=0.50\columnwidth]{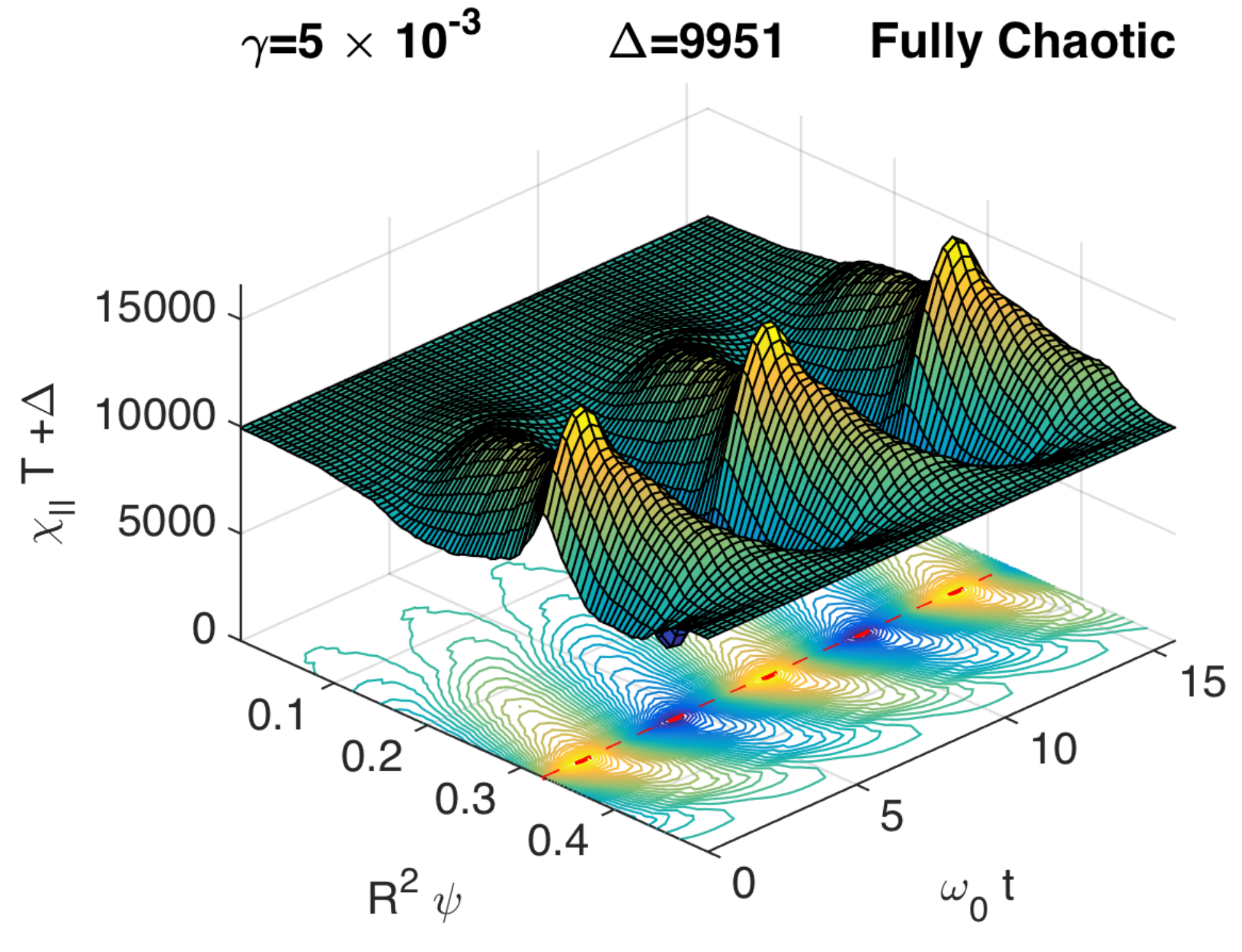}
\includegraphics[width=0.50\columnwidth]{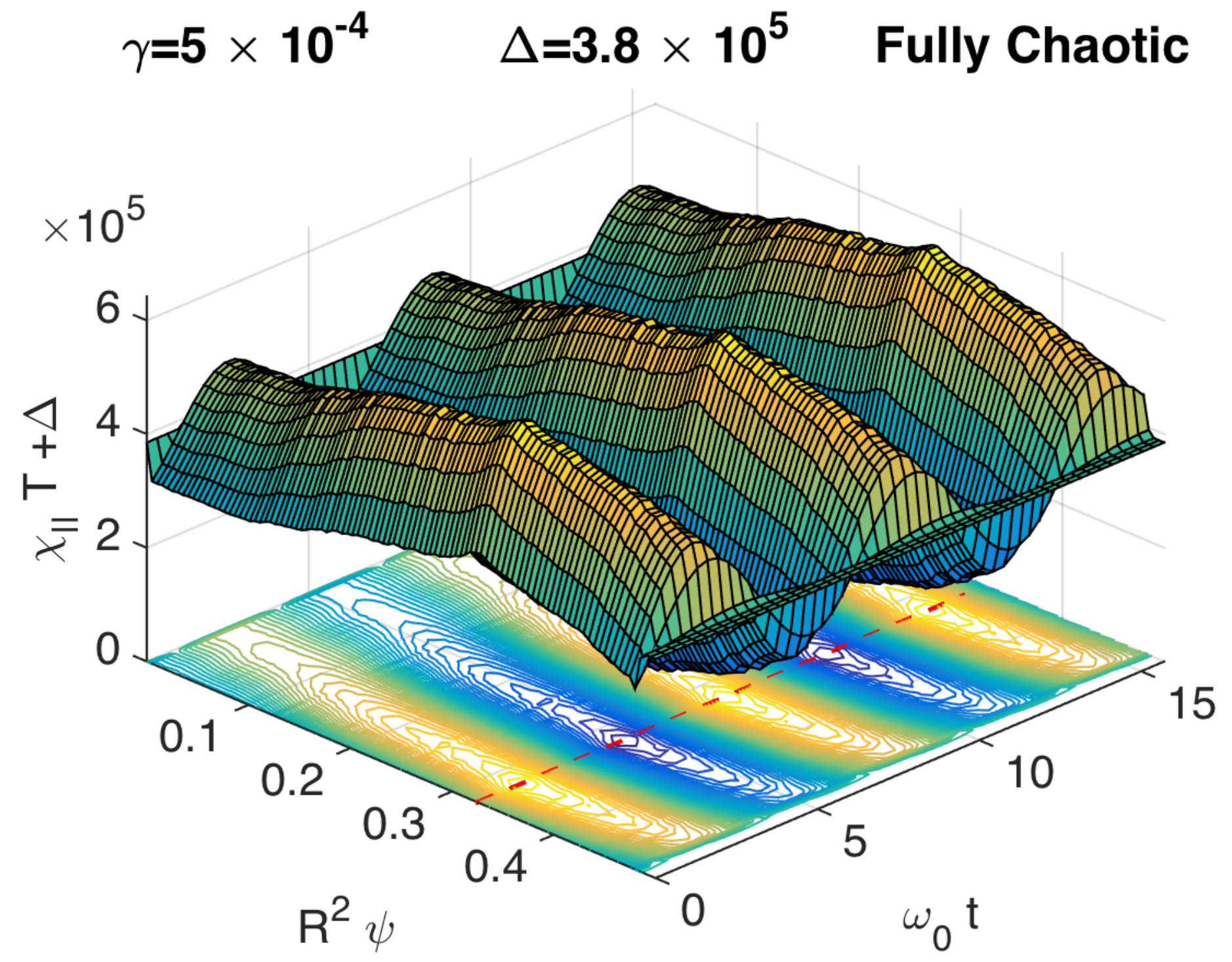}
\includegraphics[width=0.50\columnwidth]{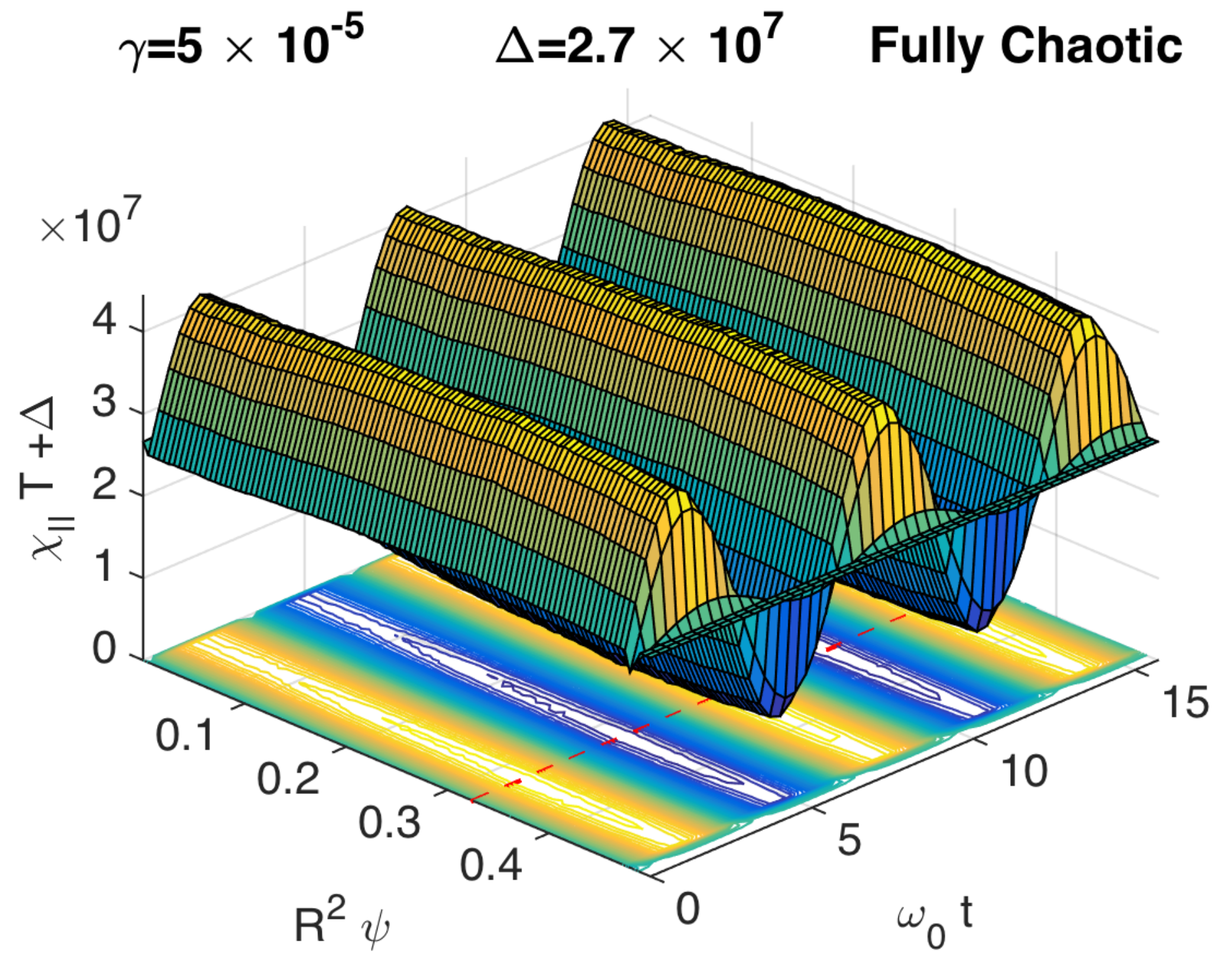}
\caption{(Color online) 
Space-time evolution of temperature modulation amplitude
(averaged over $z$ and $\theta$) for  global strong stochasticity and different values of inverse 
penetration length scale $\gamma$ in Eq.~(\ref{gamma_def}). For illustration purposes the vertical axis has been shifted by $\Delta$. The dashed red line denotes the location of the source in Eq.~(\ref{source_rt}).}
\label{T_space_time_strong_stocha}
\end{figure}

\begin{figure}
\includegraphics[width=0.45\columnwidth]{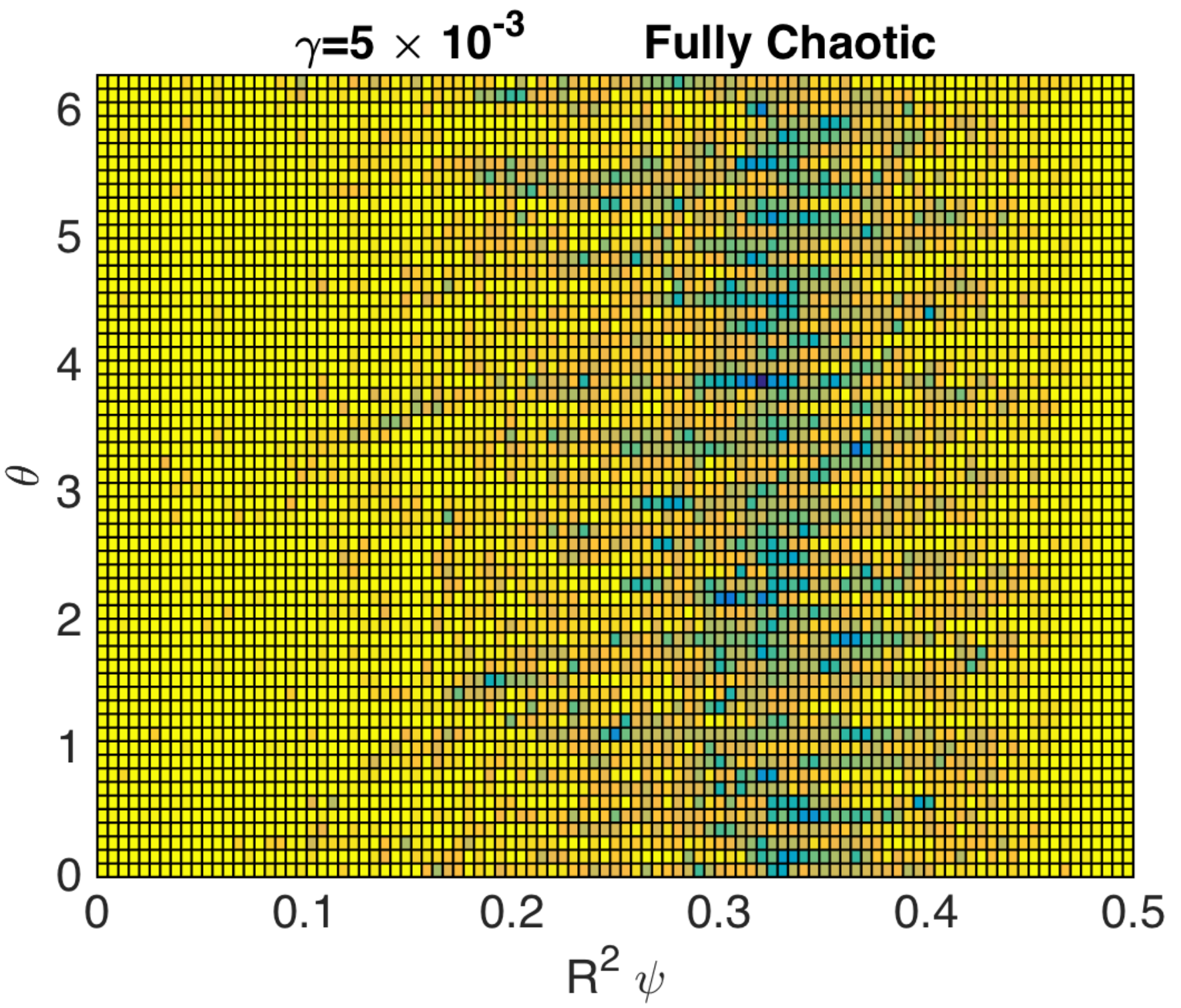}
\includegraphics[width=0.45\columnwidth]{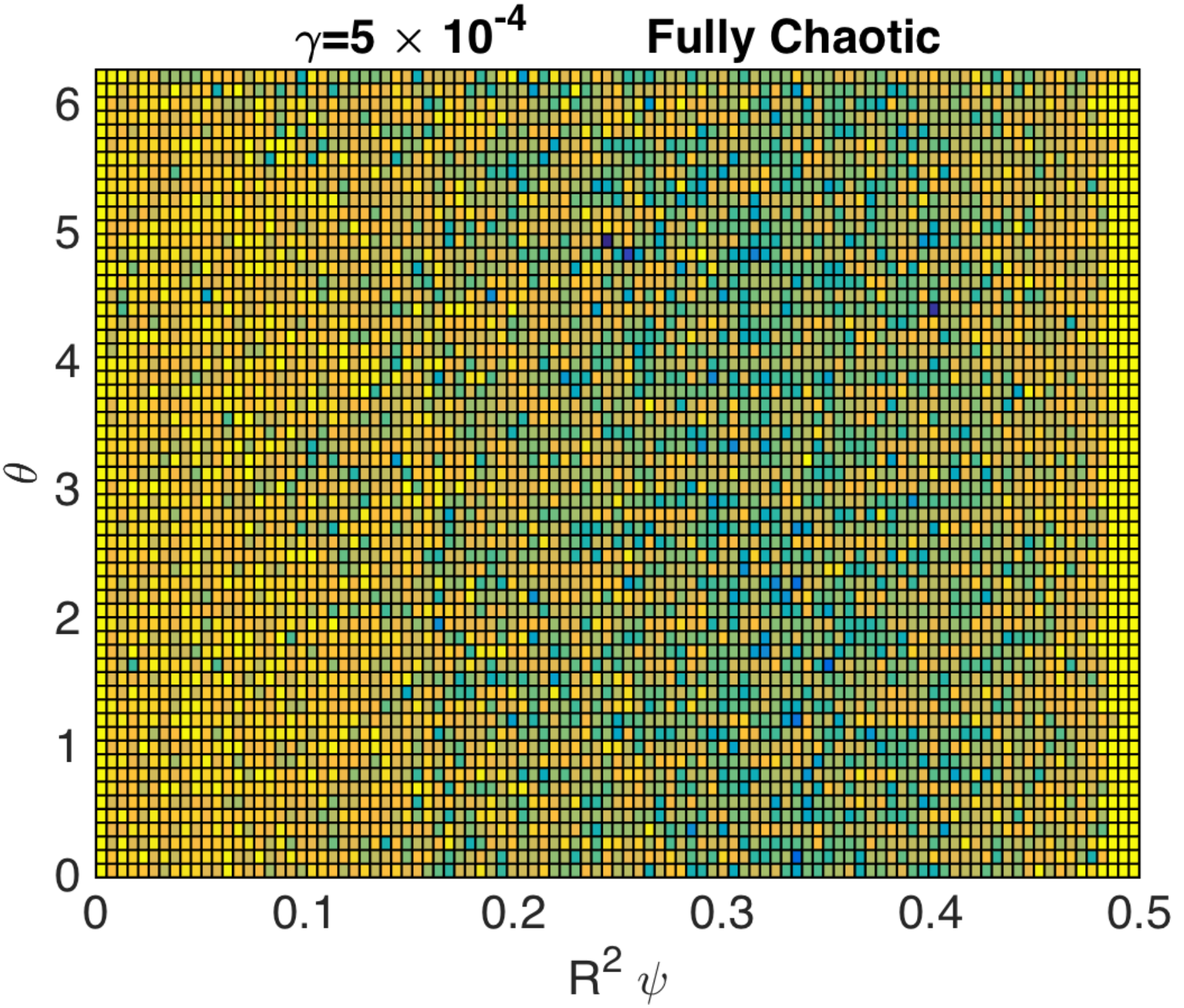}
\includegraphics[width=0.45\columnwidth]{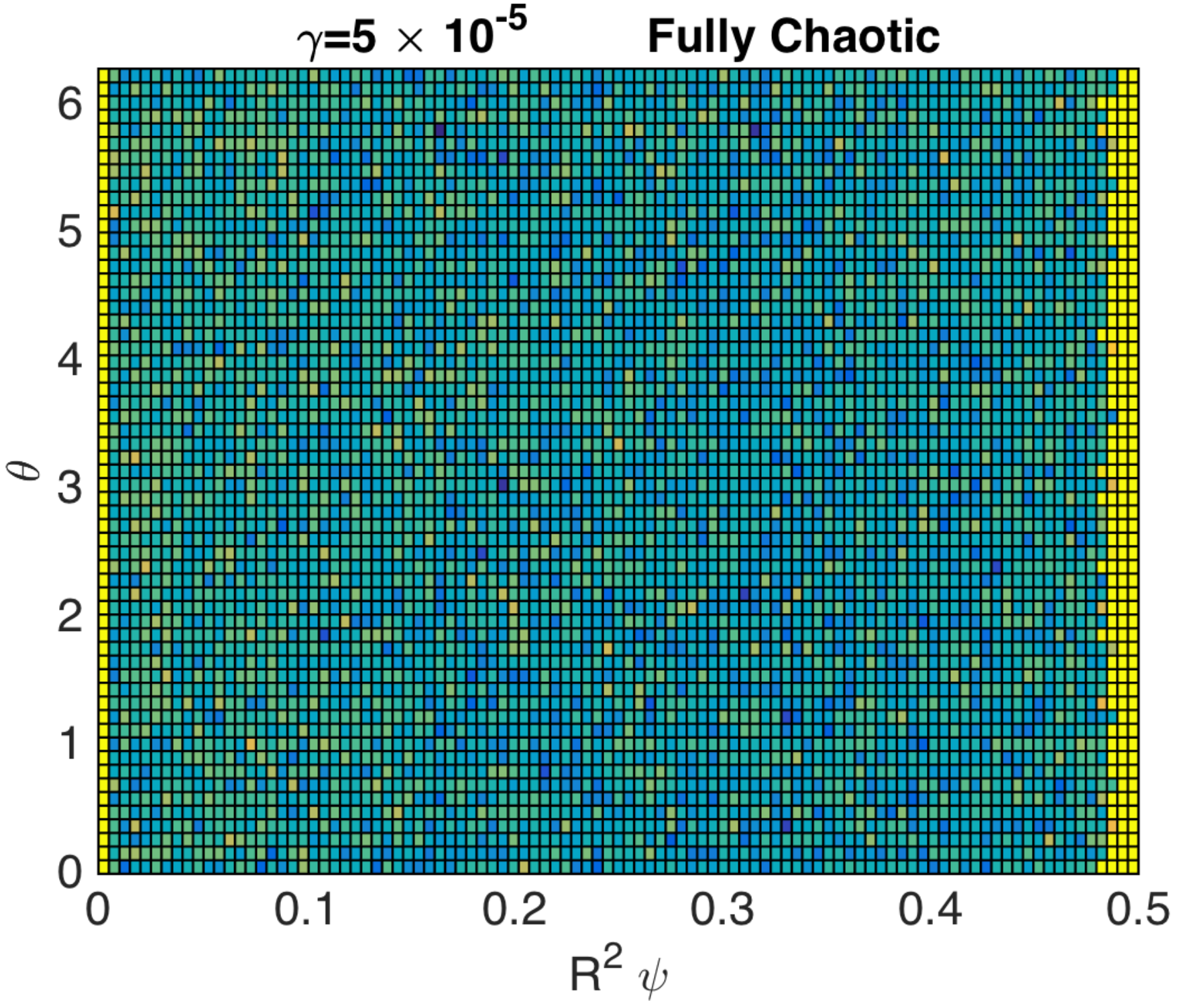}
\caption{(Color online) 
Spatial dependence in $(R^2 \psi, \theta)$ plane (at fixed $z$)  of temperature modulation amplitude
in Eq.~(\ref{rho_phi_def}) for global strong stochasticity and different values of inverse 
penetration length scale $\gamma$ in Eq.~(\ref{gamma_def}). Dark blue (light yellow) corresponds to large (small) values. 
For the quantitative ranges of amplitude values refer to Fig.~\ref{T_profiles_strong_ch}.}
\label{T_psi_theta_strong_ch}
\end{figure}

\begin{figure}
\includegraphics[width=0.45\columnwidth]{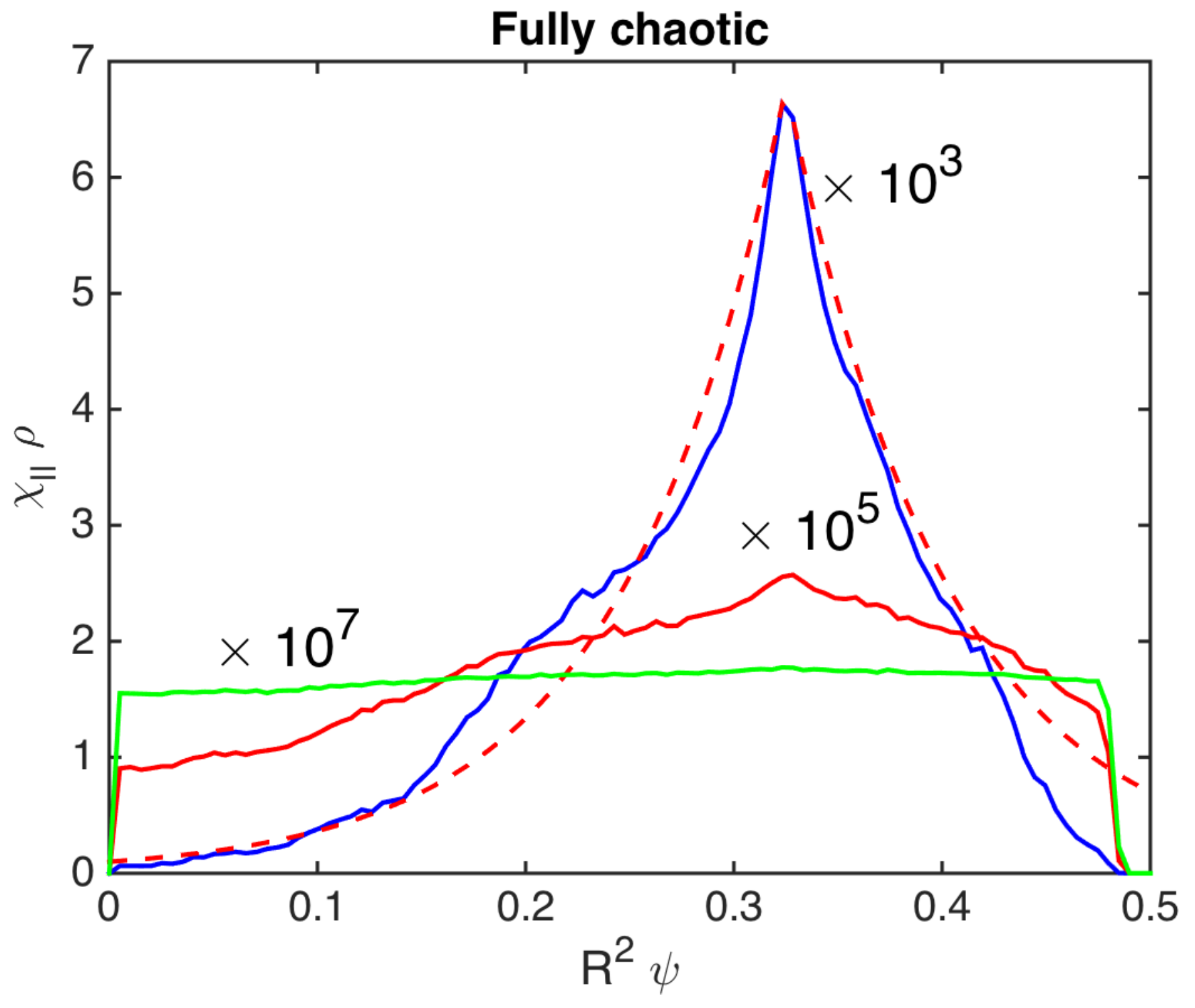}
\includegraphics[width=0.45\columnwidth]{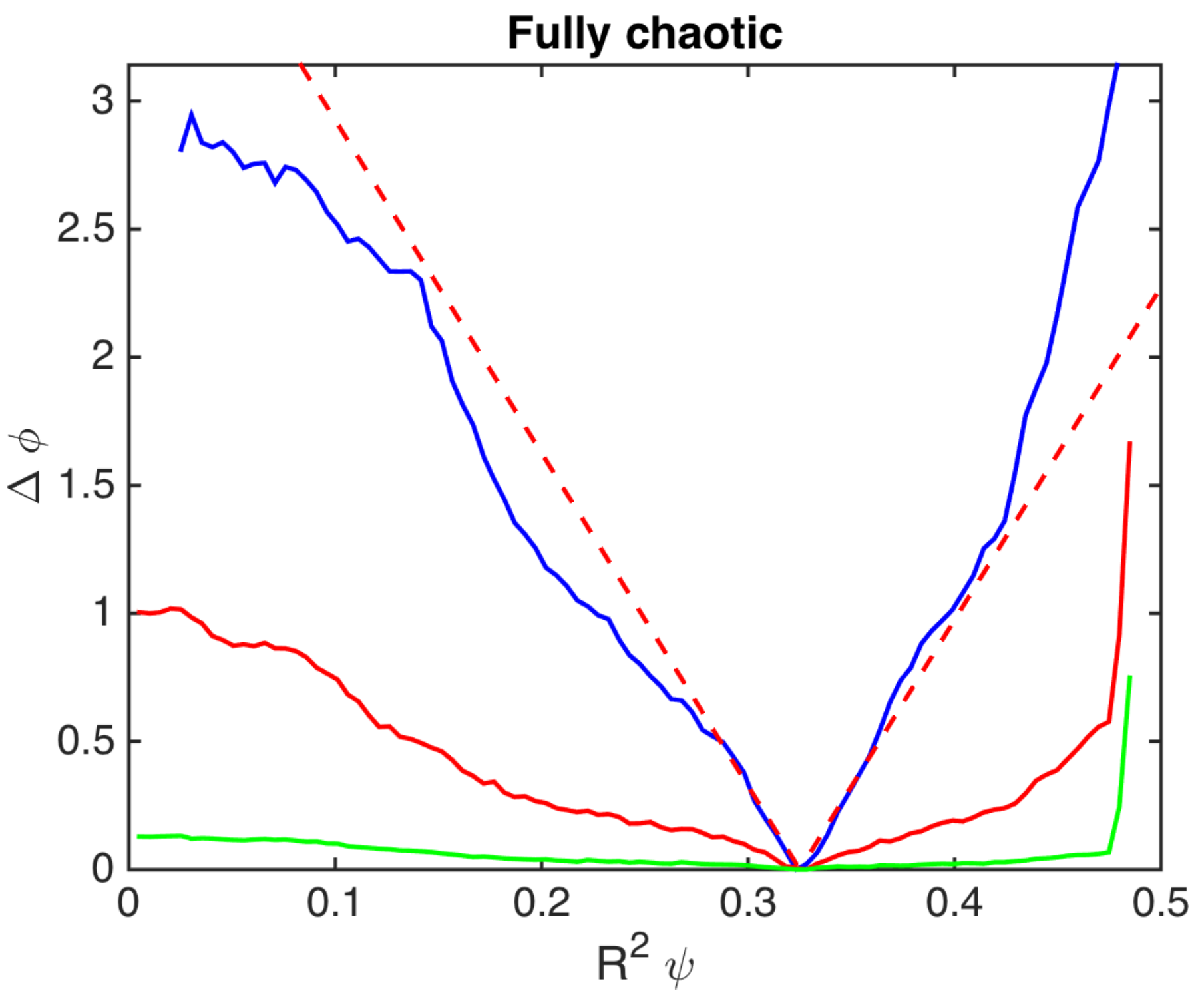}
\caption{(Color online) Heat wave amplitude, $\rho$, and relative phase, $\Delta \phi$,  
in  fully chaotic magnetic field.
In each figure the different curves correspond to different values of the inverse penetration length $\gamma$. Blue $\gamma=5 \times 10^{-3}$; Red $\gamma=5 \times 10^{-4}$; Green $\gamma=5 \times 10^{-5}$. Note the use of the pre-factors,
$\times 10^3$,  $\times 10^5$, and $\times 10^7$, multiplying $\chi_{||} \rho$ in the top panel. 
The dashed lines show the fittings of the amplitude and the phase according to the effective radial diffusion model in Eq.~(\ref{diff_fits}) with  $\gamma_{eff}=13$. }
\label{T_profiles_strong_ch}
\end{figure}

The numerical results of the propagation of modulated heat pulses in this fully stochastic field are shown in Figs.~\ref{T_space_time_strong_stocha}, \ref{T_psi_theta_strong_ch} and \ref{T_profiles_strong_ch}. As in the weakly chaotic case, a strong dependence on $\gamma$ is observed. For the smallest value  considered, $\gamma=5 \times 10^{-5}$, the weak parallel damping of the heat wave along the magnetic field, coupled with the strong stochasticity of the magnetic field, leads to an almost flat radial temperature profile and a small time delay, $\tau(\psi)=\Delta \phi /\omega_0$, resulting from a relatively fast effective radial phase-speed, $V\sim d \psi / d \tau$,  of the heat wave (see Fig.~\ref{T_profiles_strong_ch}). As shown in  Fig.~\ref{T_psi_theta_strong_ch},  this case also shows a fairly uniform temperature distribution in the $(R^2 \psi, \theta)$ plane. Note also that, compared with the other larger values of $\gamma$, the net amplitude of the temperature modulation  is considerably larger. 
As the value of $\gamma$ increases, the heat wave amplitude exhibits a stronger radial decay, and despite the fact that there are no flux surfaces, almost no heat transport is observed near the boundaries of the computational domain, $R^2 \psi \sim 0$ and $R^2 \psi \sim 0.5$, in the case $\gamma = 5 \times 10^{-3}$. 

%%%%%%%%%%%%%%%%%%%%%%%%%%%%%%%%%%%%
%%%%%%%%%%%%%%%%%%%%%%%%%%%%%%%%%%%%

%%%%%%%%%%%%%%%%%%%%%%%%%%%%%%%%%%%%
%%%%%%%%%%%%%%%%%%%%%%%%%%%%%%%%%%%%

To explore the possibility of describing the radial transport of modulated heat pulses in fully chaotic magnetic fields using a diffusive  model, consider 
the $1$-dimensional radial diffusion equation in the slab (Cartesian) approximation with a time-periodic,  spatially localized source 
\bq
\label{eff_diff}
\partial_t T  = \chi_{eff} \, \partial^2_x  T   + A \delta(x-x_0) e^{i \omega_o t}
\eq
where  $\chi_{eff}$ denotes an effective radial diffusivity.
The solution of Eq.~(\ref{eff_diff}) in an unbounded domain is
\bq
T(x,t) = {\rm Re} \frac{A }{\left(1 + i \right ) \gamma_{eff} \, \chi_{eff}} \, 
\exp 
\left[ -\gamma_{eff}  \left| x-x_0 \right | \right] 
\exp 
\left[ -i \left( \gamma_{eff} \left| x-x_0 \right |- \omega_0 t \right) \right] \, ,
\eq
where $\gamma_{eff}=\sqrt{\omega_0 / 2 \chi_{eff}}$. This implies the following radial dependence of  the amplitude and the relative (with respect to the value at $x=x_0$) phase
\bq
\label{diff_fits}
\chi_{eff}\, \rho (x) = \frac{A }{\sqrt{2}  \gamma_{eff} } \, 
\exp 
\left[ -\gamma_{eff} \left| x-x_0 \right | \right] \, , \qquad
\Delta \phi (x) = \gamma_{eff} \left| x-x_0 \right | \, .
\eq 
That is, the radial diffusive transport of a modulated heat pulse (in the slab approximation) exhibits an
exponential spatial-decay  in the amplitude and a linear increase of the time delay, which corresponds to the constant phase velocity, $V = {\rm sgn} \, \omega_0/\gamma_{eff}$, where ${\rm sgn}$ denotes the sign of $x-x_0$. 

Figure~\ref{T_profiles_strong_ch}  compares the
direct numerical simulation of the parallel heat transport equation (in the fully chaotic magnetic field with $\gamma = 5 \times 10^{-3}$)  with an exponential fit for the amplitude, with decay $-\gamma_{eff}$, and a linear fit for the phase, with slope $-\gamma_{eff}$, according to Eq.~(\ref{diff_fits}).
The relative good fitting  supports a diffusive transport description of the modulated heat pulse despite the slab approximation, the assumption of an unbounded domain, and the approximation of the narrow Gaussian profile in Eq.~(\ref{source_rt}) with a Dirac delta function. From Eq.~(\ref{gamma_def}) and 
 $\gamma_{eff}=\sqrt{\omega_0 / 2 \chi_{eff}}$
it follows that the effective radial conductivity, $\chi_{eff}$, and the parallel conductivity , $\chi_{||}$,
satisfy $\chi_{eff}/\chi_{||}=\left(\gamma_{||}/\gamma_{eff}\right)^2$. For the value used in the fitting shown in Fig.~\ref{T_profiles_strong_ch}, $\gamma_{eff}=13$, and $\gamma_{||}=5 \times 10^{-3}$, we conclude that $\chi_{eff} \sim 2 \times 10^{-7}\chi_{||}$, where the order of magnitude of the pre-factor depends on the amplitude of the magnetic field perturbation $\epsilon$.

%%%%%%%%%%%%%%%%%%%%%%%%%%%%%%%%%%%%
%%%%%%%%%%%%%%%%%%%%%%%%%%%%%%%%%%%%

It is interesting to compare the results presented here with those discussed in 
Refs.~\cite{DL_pop,diego_dan_2014} where it was concluded that,
for the same magnetic field,  the decay of pulses exhibit sub-diffusive behavior with non-Gaussian (stretched exponential) self-similar scaling.  The apparent discrepancy between these two results stems from the key role played by the modulation frequency. 
Contrary to the forced problem discussed here,  the pulse decay problem in Refs.~\cite{DL_pop,diego_dan_2014} is an initial value problem with  a sharply localized initial condition 
$T(r,t=0)$ in the absence of a source. As a result, the non-local transport processes responsible for the observed non-diffusive behavior of {\em decaying} heat pulses are ``averaged-out" in the presence of high-enough modulation frequencies (i.e., large values of $\gamma$). A clear evidence of this was originally observed in Ref.~\cite{diego_etal_2008} that compared the solutions of {\em non-local} and diffusive heat transport equations for decaying pulses and heat waves driven by power modulation. Reference~\cite{diego_etal_2008} showed that although non-locality clearly manifests in the non-diffusive evolution of decaying pulses, it does not have a significant role in the case of power modulation. In particular,  in the high-frequency regime it was observed that the solution of the non-local equation was almost indistinguishable from that of the local diffusive model. 

%%%%%%%%%%%%%%%%%%%%%%%%%%%%%%%%%%%%%%%%
\subsection{Transport in the presence of a magnetic island}
\label{island_results}
%%%%%%%%%%%%%%%%%%%%%%%%%%%%%%%%%%%%%%%%

\begin{figure}
\includegraphics[width=0.80\columnwidth]{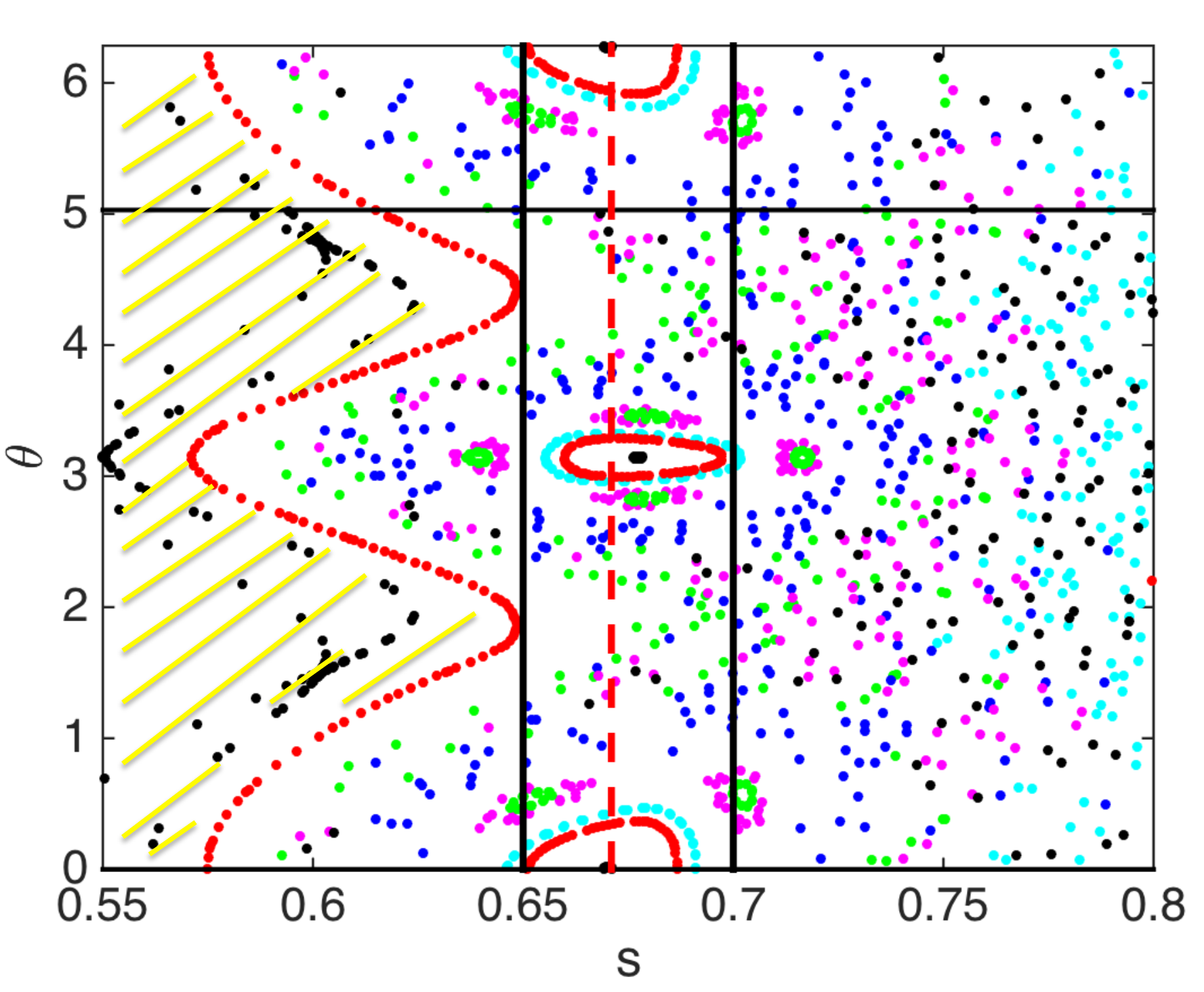}
\caption{(Color online) 
Poincare plot at the $v=0$ poloidal plane of the magnetic field used in the study of transport across a magnetic island in toroidal geometry. 
The horizontal lines at $\theta=0$ and $\theta=5$ indicate the $\theta$-cuts used in Figs.~\ref{T_space_time_island}
and \ref{T_profiles_SIESTA}. 
The dashed yellow lines denote the source, and the  red dashed line at $s =0.67$ denotes the location of the 
$(m,n)=(2,1)$ resonance. The vertical lines at $s=0.65$ and $s=0.7$  mark the location of the magnetic island. }
\label{T_Poincare_SIESTA_island}
\end{figure}

To conclude,  we discuss preliminary results on the study of the role of magnetic islands on the transport of modulated heat pulses in toroidal geometry. The magnetic field in this case was obtained using the equilibrium MHD code SIESTA \cite{hirshman_etal_2011} by imposing an initial helical perturbation with amplitude $\epsilon \sim 10^{-3}$ at the dominant $m=2$ and $n=1$ resonance surface of a DIII-D equilibrium \cite{morgan}.
The corresponding Poincare plot in Fig.~\ref{T_Poincare_SIESTA_island} shows a dominant $(m,n)=(2,1)$ island at $s \sim0.67$ surrounded by chaotic magnetic field orbits. The calculations in this case were done using coordinates $(s,u,v)$, where $u$ and $v$ are the VMEC poloidal and toroidal (cylindrical)  angles, and $s$ is the radial flux coordinate. 

%%%%%%%%%%%%%%%%%%%%%%%%%%%%%%%%%%%%
%%%%%%%%%%%%%%%%%%%%%%%%%%%%%%%%%%%%

In order to isolate the role of the magnetic island and the magnetic stochasticity, the source  was extended up to the last flux surface,  shown in red in  Fig.~\ref{T_Poincare_SIESTA_island}, by using the following model for the spatial dependence  
\bq
\label{siesta_source}
Q({\bf r})=\frac{1}{2} \left[ 1 - \tanh \left[\frac{s + f(u,v)}{h} 
\right]
\right] \, ,
\eq
where, as mentioned before, here $s$ denotes the radial flux coordinate not to be confused with the arc-length parameter, $s'$, used in the LG formula in Eq.~(\ref{LG-F}). 
In Eq.~(\ref{siesta_source}), the $\tanh$ function, with $h=0.005$, is used to localize the source to the left of the last flux surface. 
The $ f(u,v)$ is a simplified model of the poloidal and toroidal dependence of the source boundary constructed by keeping the dominant modes of the Fourier decomposition of the last flux surface.  In particular, $f=A \cos(2u+v)+B \cos(u+v)+C \sin(u+v)+D \sin(2u+v)+E$, where
$A=0.032$, $B=0.0054$, $C=-0.000024$, $D=0.00046$ and $E=0.005$. As in all the previous computations, the time dependence consisted of a single harmonic with frequency $\omega_0$, i.e. $\Lambda(t)=e^{i \omega_0 t}$. The value of $\gamma$ used in these calculations was $\gamma=5 \times 10^{-3}$. 

\begin{figure}
\includegraphics[width=0.55\columnwidth]{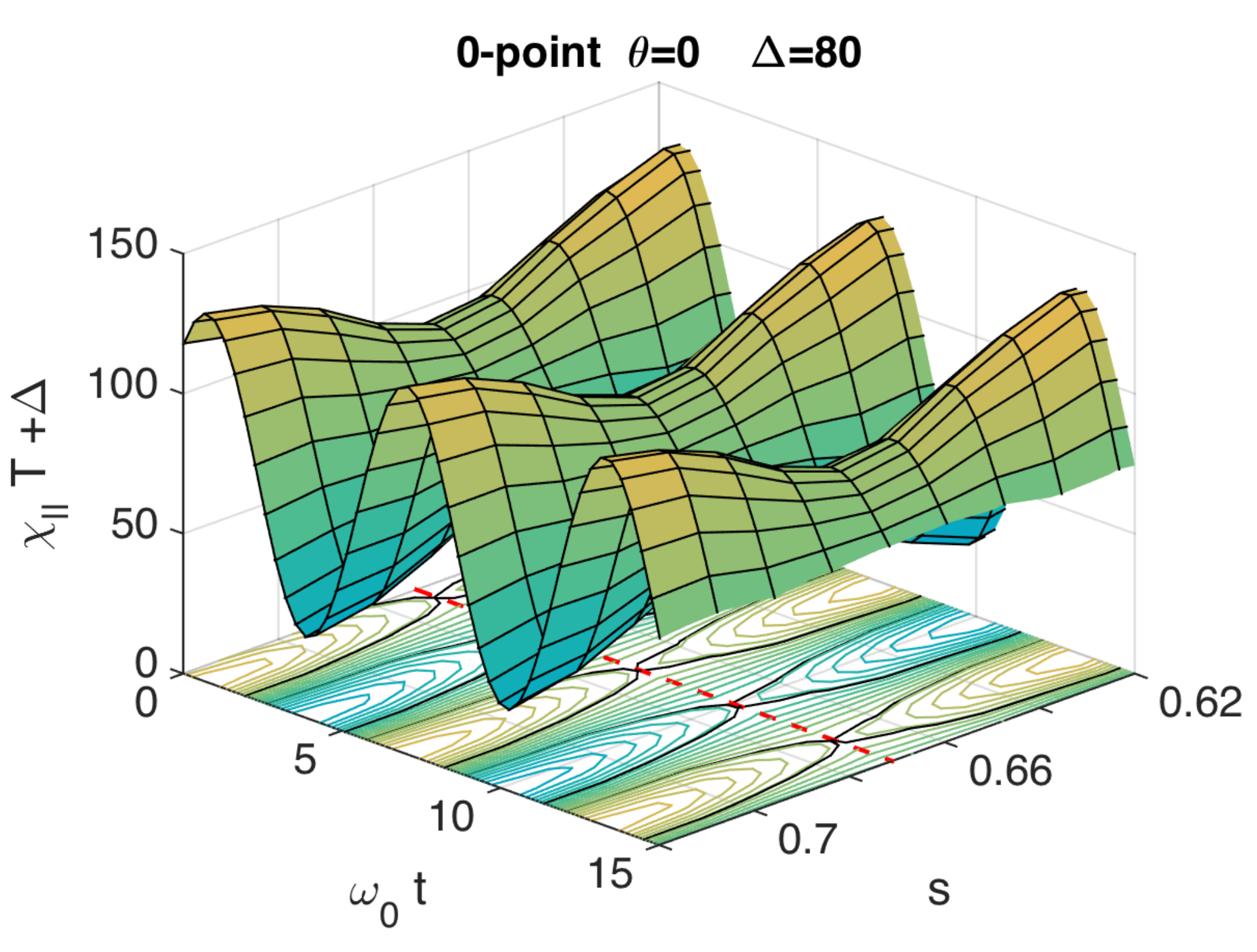}
\includegraphics[width=0.55\columnwidth]{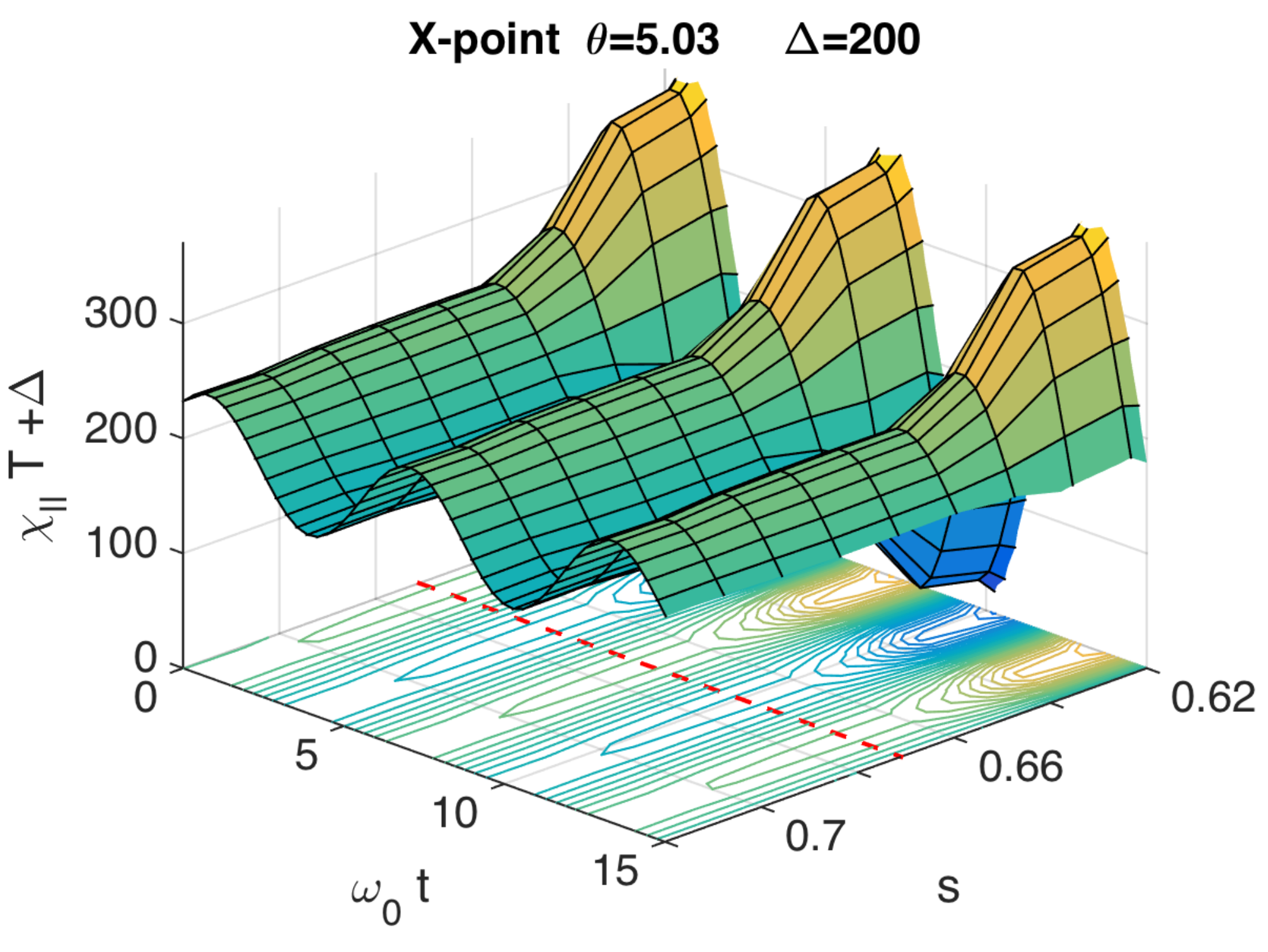}
\caption{(Color online) 
Space-time evolution of temperature modulation amplitude in the presence of a magnetic island with
$\gamma=5 \times 10^{-3}$.
The plots show the amplitude as function of radius, $s$, at  the $v=0$ poloidal plane and a fixed value of $\theta$. 
The top (bottom) panel corresponds to a cut across the O-point, $\theta=0$ (X-point, $\theta=5$) in the Poincare section shown in Fig.~\ref{T_Poincare_SIESTA_island} . The dashed red line denotes the radial location  of the magnetic island.  
For illustration purposes the vertical axes have been shifted by $\Delta$.
}
\label{T_space_time_island}
\end{figure}

To explore the role of the magnetic island topology on transport, rather than taking the toroidal and poloidal averages of the temperature (as done in the previous calculations), here we plot $T$ along different cuts of the poloidal angle $\theta$ at the fixed toroidal angle $v=0$. 
Doing this allows us to study directly the role of the O and the X points of the magnetic island on transport.  
Figure~\ref{T_space_time_island} shows the time evolution at $v=0$ of the radial dependence  of the  temperature at $\theta=0$ and $\theta=5$ which, according to Fig.~\ref{T_Poincare_SIESTA_island}, correspond to the O-point and the X-point of the magnetic island. 
It is observed that, whereas the temperature along the 
X-point exhibits a monotonic spatial dependence, the temperature along the 
O-point exhibits a non-monotonic spatial dependence. In particular, the space-time temperature evolution 
across the O-point  of the magnetic island shows alternating  saddle points that project on the $(s,\omega_0 t)$ contour plot as X-points at $s=0.67$ (which, as seen in Fig.~\ref{T_Poincare_SIESTA_island} is the location of the magnetic island)  and at $\omega_0 t=\pi/4+n \pi$, where $n=1,\,2,\,\ldots$. On the other hand, the space-time temperature evolution across the X-point of the magnetic island
exhibits a monotonic penetration of the heat wave with an overall larger amplitude.  

%%%%%%%%%%%%%%%%%%%%%%%%%%%%%%%%%%%%
%%%%%%%%%%%%%%%%%%%%%%%%%%%%%%%%%%%%

\begin{figure}
\includegraphics[width=0.45\columnwidth]{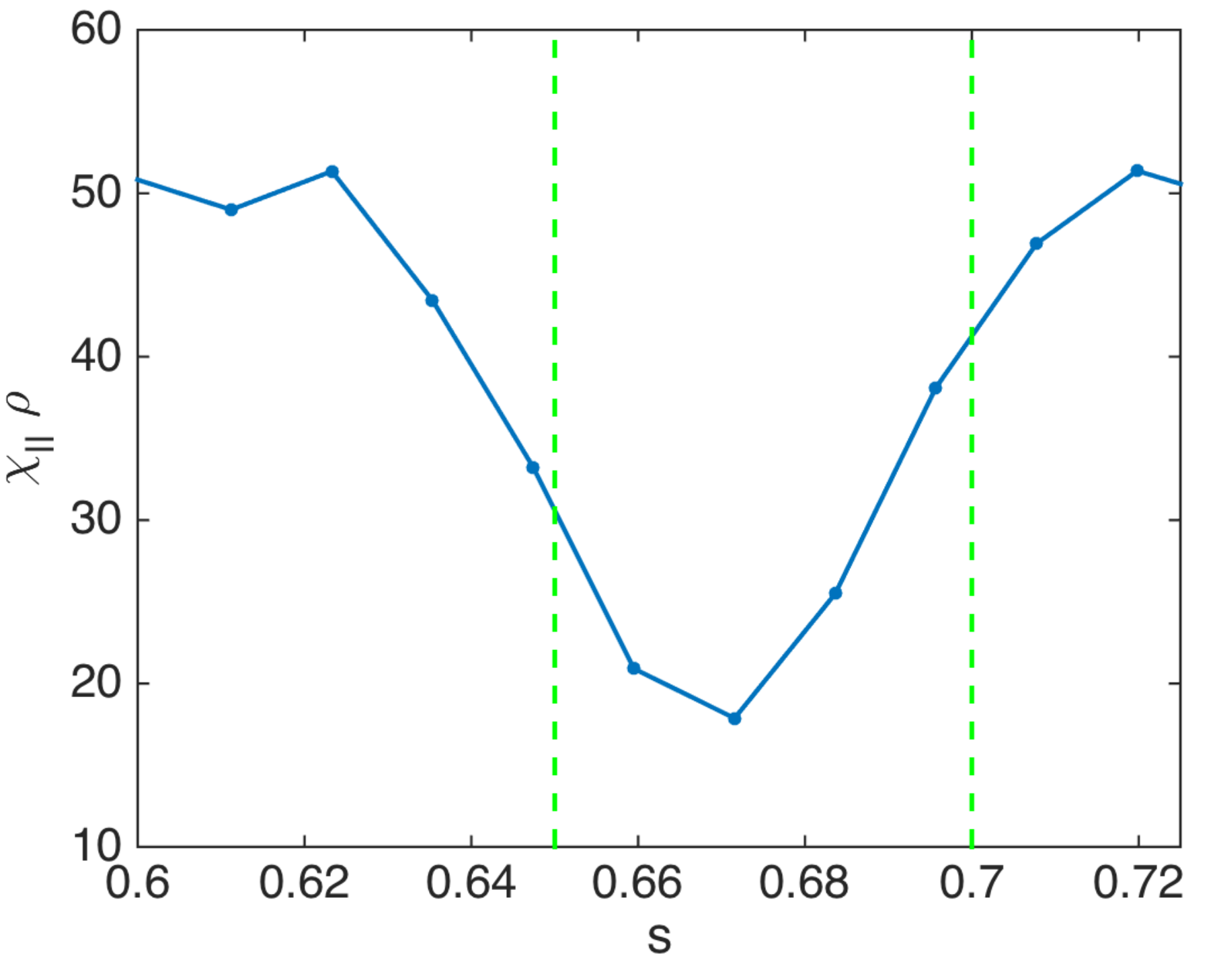}
\includegraphics[width=0.45\columnwidth]{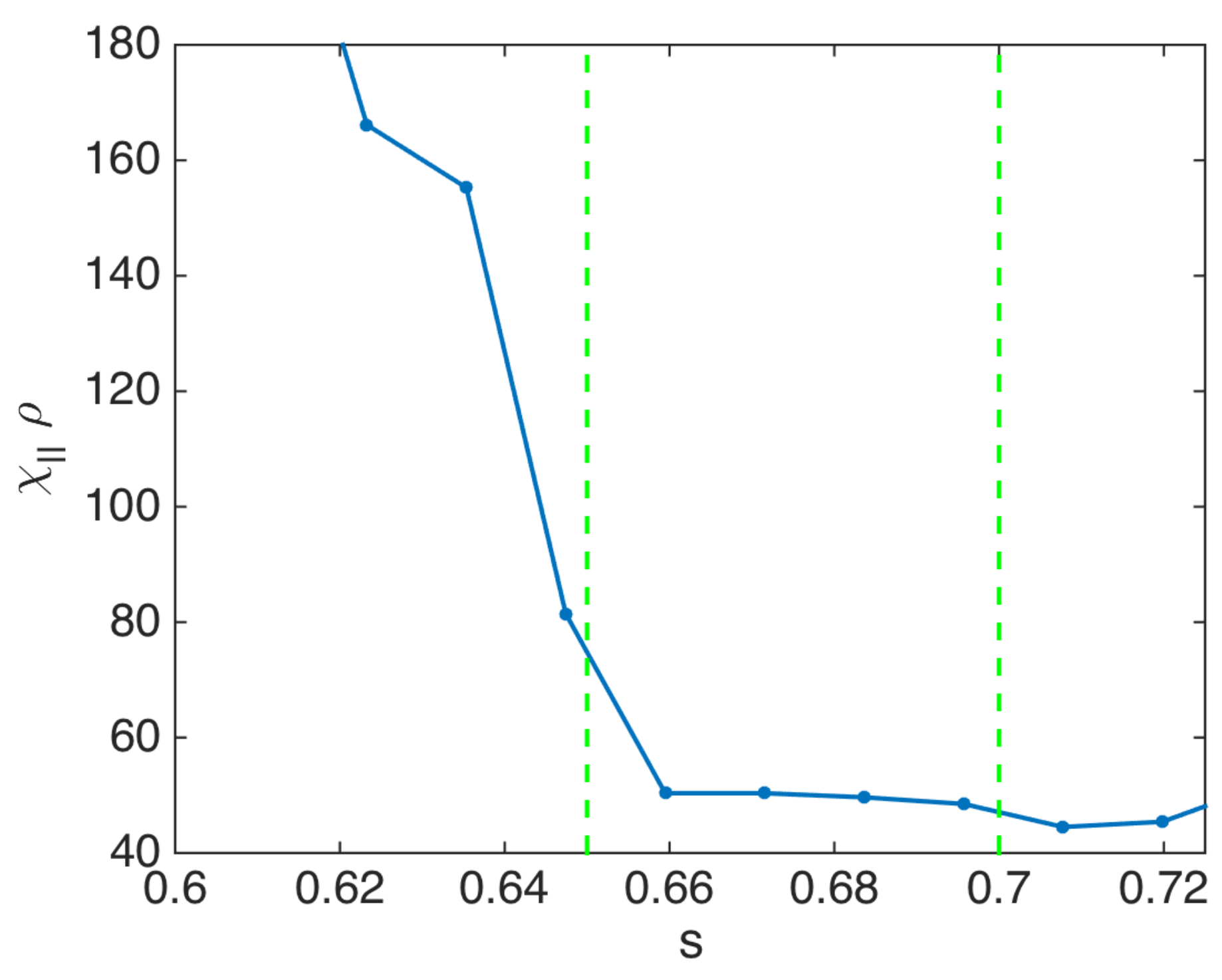}
\caption{(Color online) Radial dependence of temperature modulation amplitude, $\rho$, at  the $v=0$ poloidal plane and fixed $\theta$.
The left (right) panel corresponds to  a $\theta=0$ ($\theta=5$) cut across the O-point (X-point). The vertical lines at $s=0.65$ and $s=0.7$  mark the location of the magnetic island  in the Poincare plot in 
Fig.~\ref{T_Poincare_SIESTA_island}.}
\label{T_profiles_SIESTA}
\end{figure}

Figure~\ref{T_profiles_SIESTA} shows the spatial dependence of the amplitude. As expected, in the case of the X-point a flat region is observed between $s=0.65$ and $s=0.70$.
Comparing Fig.~\ref{T_Poincare_SIESTA_island} and \ref{T_profiles_SIESTA} it is observed that the region, $s \sim (0.65, 0.70)$, where the amplitude exhibits its local minimum in the case of the O-point, gives a good estimate of the width, $W \sim 0.05$, of the magnetic island. 
There is qualitative agreement between these results and some of the recent experimental results reported in Ref.~\cite{Ida_etal_2015} where the relationship between the magnetic field topology and the propagation of heat pulses was studied using electron cyclotron heating (ECH) power modulation. For example, the spatio-temporal evolution of the relative modulation amplitude of the electron temperature in Fig.~4 of Ref.~\cite{Ida_etal_2015} is qualitatively similar to the contour plots in Fig.~\ref{T_space_time_island}.  Also, the radial dependence of the modulation amplitude across the X and the O points in Fig.~5 of Ref.~\cite{Ida_etal_2015}  is qualitative similar to the one observed in Fig.~\ref{T_profiles_SIESTA}. 

%%%%%%%%%%%%%%%%%%%%%%%%%%%%%%%%%%%%
%%%%%%%%%%%%%%%%%%%%%%%%%%%%%%%%%%%%

%%%%%%%%%%%%%%%%%%%%%%%%%%%%%%%%%%%%%%%%
\section{Summary and conclusions}
\label{conclusions}
%%%%%%%%%%%%%%%%%%%%%%%%%%%%%%%%%%%%%%%%

We have presented a study of the propagation of modulated heat pulses in three-dimensional chaotic magnetic fields with different levels of stochasticity. The numerical method was based  on the Fourier formulation of the Lagrangian-Green's function method that provides an accurate and efficient technique for the solution of the parallel heat transport equation in the presence of power modulation. In particular, the numerical computation of the harmonic time dependence of temperature at 
each point in space is reduced to the integration of six ordinary differential equations (odes) determining the forward and backwards  3-dimensional magnetic field orbits and two odes determining the amplitude of the temperature modulation. 
The two key parameters in the simulations are the strength of the magnetic field perturbation, $\epsilon$, and the inverse penetration length scale, $\gamma=\sqrt{\omega_0/2 \chi_{||}}$, where $\omega_0$ is the power modulation frequency and $\chi_{||}$ the parallel thermal conductivity. 

In all the numerical simulations presented, there are no magnetic flux surfaces in the computational domain of interest. That is, there is no magnetic field line confinement. However, the topology of the magnetic field is far from trivial and it exhibits partial transport barriers known as Cantori. These structures, which were originally discovered in studies of Hamiltonian chaos, appear as remnants of broken magnetic flux surfaces and their intricate structure consisting of a fractal distribution of gaps gives rise to anomalous transport.  To characterize the magnetic field structure we use the magnetic field connection length that measures the number of iteration in the Poincare map needed to transition from one part of the domain to another.  A clear relationship was observed between the radial  gradient of the magnetic field connection length and the location of partial barriers for all the values of  $\epsilon$ considered. 

Depending on the conditions, heat transport can ``leak" through partial barriers. 
For a given level of magnetic field line stochasticity,  parallel heat transport is strongly damped whenever $\omega_0$ is large or $\chi_{||}$ is small, that is for  $\gamma \gg1 $.  As a result, in this case the amplitude of the heat wave vanishes and its phase speed slows down to a halt at the location of the partial barrier. 
On the other hand, in the limit of small $\gamma$, that is, for small  $\omega_0$  or large $\chi_{||}$,  parallel heat transport is largely unimpeded and global radial heat transport is observed accompanied by a significant time delay, 
$\tau=\Delta \phi/\omega_0$,  across the partial barrier. One of the main conclusions implied by this study is that the connection between magnetic field line stochasticity and heat transport is far from trivial. In particular, the lack of magnetic 
field line confinement does not necessarily imply lack of transport barriers in the propagation of modulated heat pulses.
 Cantori and high order magnetic islands not only impact the radial propagation,  these structures also manifest in the poloidal dependence of the amplitude of the heat waves that resembles the poloidal dependence of the magnetic field connection length. The relationship between the flux of field lines across Cantori and the heat flux computed by the LG method 
is an interesting problem that we plan to address in a future publication. This problem can be studied by first computing the radial heat flux outside the source by using Eq.~(35) in Ref.~[4] with $\langle T \rangle$ given by the LG solution in 
Eq.~(\ref{rho_phi_def}) of the present paper.  This flux should then be compared with the field lines flux across Cantori computed from the difference of the Lagrangian actions of properly chosen orbits as discussed in Refs.~\cite{mckay,hudson_2006}.  

We also considered heat pulse propagation in the case of very strong magnetic field stochasticity lacking magnetic islands and partial barriers. We showed that in this extreme stochastic regime transport is diffusive for high enough  modulation frequencies (i.e., large values of $\gamma$). In particular, for the case studied, is was conclude that the effective radial diffusivity is of the order $\chi_{eff} \sim 10^{-7} \chi_{||}$. However, care must be observed when attempting to extend this conclusion beyond the specific parameter regimes and type of sources considered. In particular,  in the case of freely decaying pulses (i.e. in the absence of sources) conclusive evidence of non-diffusive effective radial transport has been documented before in the literature. 

Finally, motivated by  recent experimental studies in DIII-D and LHD we presented preliminary results on modulated heat pulse propagation across magnetic islands. The geometry used in this calculation was toroidal and the magnetic field was obtained using the MHD equilibrium code SIESTA. We showed that the magnetic island, in particular the elliptic (O) and hyperbolic (X) points have a direct impact on the spatio-temporal evolution of modulated heat pulses. 
Qualitative agreement was observed with the experimental results. In particular, whereas the temperature along the 
X-point exhibits a monotonic spatial dependence, the temperature along the 
O-point exhibits a non-monotonic spatial dependence. The region where the amplitude exhibits its local minimum across the O-point  was shown to give a good estimate of the width of the magnetic island.
%%%%%%%%%%%%%%%%%%%%%%%%%%%%%%%%%%%%%%%%

 % %%%%%%%%%%%%%%%%%%%%%%%%%
 \section{Acknowledgments}
 % %%%%%%%%%%%%%%%%%%%%%%%%%
 
We thank Morgan Shafer for kindly proving the SIESTA magnetic field data. 
This work was sponsored by the Office of Fusion Energy Sciences 
of the US Department of 
Energy at Oak Ridge National Laboratory, managed by UT-Battelle, LLC, 
for the U.S.Department of Energy under contract DE-AC05-00OR22725.

\end{document}